\documentclass[12pt,twoside]{article}

\usepackage{amsmath,amssymb,bbm,feynmf,epsfig,rotating,subfigure,here,citesort}

\begin{document}

\title{
\vspace{-0.6cm}
{\sf \normalsize \rightline{HD-THEP-00-65}}
\vspace*{0.8cm}
Inelastic diffractive scattering in nonperturbative QCD}
\author{{\sc H.G. Dosch}\footnote{\tt H.Dosch@ThPhys.Uni-Heidelberg.DE}, {\sc O. Nachtmann}\footnote{\tt O.Nachtmann@ThPhys.Uni-Heidelberg.DE}, {\sc T. Paulus}\footnote{\tt T.Paulus@ThPhys.Uni-Heidelberg.DE}, {\sc S. Weinstock}\footnote{{\tt S.Weinstock@ThPhys.Uni-Heidelberg.DE}\newline \newline Supported by German Bundesministerium f\"ur Bildung und Forschung (BMBF), Contract no. 05 HT 9VHA3 and by Deutsche Forschungsgemeinschaft (DFG), Grant no. GRK 216/1/00}}
\date{}

\maketitle

\begin{center}
{\em Institut f\"ur Theoretische Physik,\\Universit\"at Heidelberg,\\Philosophenweg 16,\\D-69120 Heidelberg, Germany}
\end{center}

\bigskip

\begin{abstract}
We examine diffractive proton-proton scattering $p \: p \to p \: X$ and photo- and electroproduction of $\rho^0$ mesons $\gamma^{(*)} \: p \to \rho^0 \: X$, where $X$ denotes a proton or a final state, into which the proton can go by diffractive dissociation.
Using a functional integral approach we derive the scattering amplitudes, which are governed by the expectation values of light-like Wegner-Wilson loops,
which are then evaluated using the model of the stochastic vacuum.
For the proton, we assume a quark-diquark structure. From the scattering amplitudes we calculate total and differential cross sections for high centre of mass energy and small momentum transfer and compare with experiments. Furthermore we calculate isovector form factors for the proton and the pion within the same model.
\end{abstract}

\newpage
\section{Introduction}

In this article we study inelastic diffractive scattering at high centre of mass (c.m.) energies $\sqrt{s} \gtrsim 20 \mbox{ GeV}$ and small momentum transfer squared $|t| \lesssim 1 \mbox{ GeV}^2$.
The low momentum transfer implies that one has to apply nonperturbative methods to investigate these processes.\\
Our numerical analysis will be focused on two types of reactions: hadron-hadron scattering (especially $p \: p \to p \: X$) and photo- and electroproduction of a $\rho^0$ meson $\gamma^{(*)} \, p \to \rho^0 \, X$.\\
The experimentally observed increase of total cross sections for hadronic reactions with the c.m. energy~\cite{PDG}, starting at about $\sqrt{s}=10 \mbox{ GeV}$, has been described phenomenologically by Donnachie and Landshoff~\cite{DL} in the context of Regge theory. 
The rise of the total hadronic cross sections can be described by a pomeron with an intercept slightly larger than one~\cite{PDG,DL}.
For inelastic diffraction, the pomeron-photon analogy was applied in~\cite{DL2} to relate the cross section of these reactions in a quantitative way to the structure functions of deep inelastic electron-proton scattering. A number of other methods to deal with inelastic diffractive reactions have been proposed (for reviews see~\cite{Forshaw,Hebecker,Goulianos}).\\
A description of soft hadronic high energy reactions, starting from a microscopic level, was developed in~\cite{LN} where in the case of an abelian gluon model the pomeron properties were related to nonperturbative aspects of the vacuum like the gluon condensate introduced by Shifman, Vainshtein and Zakharov~\cite{SVZ}. These methods were generalised to QCD in~\cite{Nachtmann}. The quantity governing the scattering amplitude was found to be a correlation function of two light-like Wegner-Wilson loops~\cite{MSVmink,DFK}. These correlation functions are evaluated using the model of the stochastic vacuum (MSV)~\cite{MSV} as formulated in Minkowski space in~\cite{MSVmink,DFK,DRio}. This method has been applied to various reactions, for example exclusive vector meson production~\cite{DGKP,DGP98,KDP99}, elastic hadron-hadron scattering~\cite{Berger}, and photo- and electroproduction of pseudoscalar and tensor mesons~\cite{Berger2,Berger3}.\\
\\
In this paper we will extend the model to the description of inelastic diffractive hadron-hadron scattering and vector meson photo- and electroproduction. The hadronic scattering amplitudes as derived in~\cite{Nachtmann,MSVmink,DFK,DRio,DGKP,DGP98,KDP99,Berger,Berger2,Berger3,Nachtmannreport,DFK2} and a brief summary of the basic features of the MSV in its Minkowskian formulation will be given in the chapters~\ref{scat_sect} and~\ref{eval_sect}. Previous results for $\mbox{d}\sigma / \mbox{d}t$ in the case of elastic scattering~\cite{Berger} are also reviewed briefly. To obtain results for single diffractive dissociation in proton-proton scattering and $\rho^0$-photo- and electroproduction, the loop-loop correlation functions will be calculated applying two approaches. Moreover we use two different models for the diffractive final state $X$. In this work, the constituent configuration of baryons is assumed to be of the quark-diquark type for the reasons given in~\cite{RDosch}. Then baryons act as colour dipoles like mesons. The numerical results of all reactions considered here are presented and compared to experimental data. Finally we calculate the isovector form factors for the proton and pion within our model in chapter~\ref{ff_sect}. Our conclusions and a summary are given in chapter~\ref{concl_sect}.

\bigskip
\section{The scattering amplitudes} \label{scat_sect}

In this section we present our basic formulae for hadron-hadron scattering and photo- and electroproduction. We treat exclusive as well as inclusive processes, where in the inclusive case only diffractive dissociation of one of the participating hadrons is considered.

\subsection{Hadron-hadron scattering}\label{scat_amp_hh_sect}

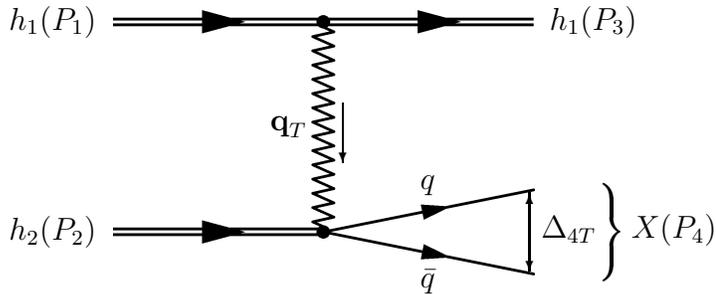
\begin{figure}[h]
\begin{center}\unitlength.8mm
\begin{picture}(80,40)
\begin{fmffile}{process}
\put(0,0){
  \begin{fmfgraph*}(70,35)
    \fmfstraight
    \fmfleft{l1,l2}\fmfright{r1,r2,r3,r4}
    \fmfforce{0.5w,0h}{r2}
    \fmfforce{1w,-.2h}{r1}
    \fmfforce{1w,.2h}{r3}
    \fmf{dbl_plain_arrow}{l1,r2}
    \fmf{dbl_plain_arrow}{l2,v,r4}
    \fmffreeze
    \fmf{zigzag,label=${\bf q}_{T}$,l.side=left}{r2,v}
    \fmf{fermion,label=$\bar q$,l.side=right}{r2,r1}
    \fmf{fermion,label=$q$,l.side=left}{r2,r3}
    \fmfv{label=$h_1(P_1)$,l.angle=180}{l2}
    \fmfv{label=$h_1(P_3)$,l.angle=0}{r4}  
    \fmfv{label=$h_2(P_2)$,l.angle=180}{l1}
    \fmfdot{r2}
    \fmfdot{v}
  \end{fmfgraph*}}
\put(73,-1){${\bf \Delta}_{4T}$}
\put(71,0){\vector(0,1){6.5}}
\put(71,0){\vector(0,-1){6.5}}
\put(40,21.5){\vector(0,-1){10}}
\begin{rotate}{90}
  \put(-8,-83){$\underbrace{\hspace{1.28cm}}$}
\end{rotate}
\put(88,-1){$X(P_4)$}
\end{fmffile}
\end{picture}
\end{center}
\vspace*{.3cm}
\caption{{\em $h_1 + h_2 \to h_1 + X$}}\label{hh_scat_pic}
\end{figure}

Consider the reaction 
\begin{equation}\label{hh_react}
h_1(P_1) + h_2(P_2) \to h_1(P_3) + X(P_4),
\end{equation}
where $h_1$ and $h_2$ are hadrons, $X$ is again $h_2$ or a diffractive excitation of $h_2$ and $P_i$ denote the four-momenta (Fig.~\ref{hh_scat_pic}).
The hadrons $h_1, \, h_2$ are modeled as quark-antiquark and quark-diquark wave packets for mesons and baryons, respectively. For the wave functions we have chosen a Bauer-Stech-Wirbel ansatz~\cite{BWS}.
The diffractive final state $X$ is modeled by a $q \bar q$-pair (or quark-diquark pair) in a colour singlet state. We then use two approaches. In the first method we use free plane waves for the quark and antiquark. Integration over all allowed values in phase space and the closure relation then yield all possible diffractive final states $X$, where the case of elastic scattering is also included.
The second ansatz, applied to confirm the results of the first method and to gain additional insight into the structure of the calculated differential cross sections, uses the wave functions of a two-dimensional harmonic oscillator where the ground state corresponds to hadron $h_2$ and the excited states to the diffractive excitations of $h_2$. Since these eigenfunctions form a basis, the calculation of cross sections can be performed as follows: first the cross section for one specific excited state with definite quantum numbers $n,m$ is calculated and then the sum over all excited states is taken to get the inelastic inclusive diffractive cross section.

In the framework of the model presented  in~\cite{DFK,Nachtmannreport} we obtain the scattering amplitude for reaction~(\ref{hh_react}) as
\begin{eqnarray} \label{scat_amp}
  \mathcal{S}_{fi} &=& \delta_{fi}+i(2 \pi)^4 \delta(P_3+P_4-P_1-P_2) \mathcal{T}_{fi}, \nonumber \\
  \mathcal{T}_{fi} &=& (2is) \int \mbox{d}^2b_T \, e^{i{\bf q}_{T} \cdot {\bf b}_T} \hat J_{\mbox{\scriptsize{diff}}}.
\end{eqnarray}
Here $\hat J_{\mbox{\scriptsize{diff}}}$ is the diffractive profile function for which we get two expressions depending on which description of the diffractive final state is used in the calculation. For the plane wave description we obtain
\begin{eqnarray}
  \hat J_{\mbox{\scriptsize{diff}}}^{\mbox{\scriptsize{p.wave}}}({\bf b}_T,z^\prime) &=& - \int \mbox{d}^2x_T \, \mbox{d}^2y_T \int_0^1 \mbox{d}z \; w_{31}({\bf x}_T,z) \, \sqrt{2 \pi} \sqrt{2 z^\prime (1-z^\prime)} \, e^{-i {\bf \Delta}_{4T} \cdot {\bf y}_T} \, \varphi_2({\bf y}_T,z^\prime) \nonumber \\ 
  & & \Bigl \langle \mathcal{W}_+(\frac{1}{2}{\bf b}_T+(\frac{1}{2}-z){\bf x}_T,{\bf x}_T) \mathcal{W}_-(-\frac{1}{2}{\bf b}_T+(\frac{1}{2}-z^\prime){\bf y}_T,{\bf y}_T)-\mathbbm{1}\Bigr \rangle_G, \label{J_wave}
\end{eqnarray}
where ${\bf \Delta}_{4T}$ is the relative transverse momentum between the quark and the antiquark (or diquark) of $X$ (see Appendix~\ref{App_A_sect}).
For the oscillator description we obtain
\begin{eqnarray}
  \hat J_{\mbox{\scriptsize{diff}}}^{\mbox{\scriptsize{2d osc.}}}({\bf b}_T) &=& - \int \mbox{d}^2x_T \, \mbox{d}^2y_T \int_0^1 \mbox{d}z \int_0^1 \mbox{d}z^\prime \; w_{31}({\bf x}_T,z) \, X^{n,m}({\bf y}_T,z^\prime) \, \varphi_2({\bf y}_T,z^\prime) \nonumber \\ 
  & & \Bigl \langle \mathcal{W}_+(\frac{1}{2}{\bf b}_T+(\frac{1}{2}-z){\bf x}_T,{\bf x}_T) \mathcal{W}_-(-\frac{1}{2}{\bf b}_T+(\frac{1}{2}-z^\prime){\bf y}_T,{\bf y}_T)-\mathbbm{1}\Bigr \rangle_G. \label{J_osc}
\end{eqnarray}

Here $X^{n,m}({\bf y}_T,z^\prime)$ stands for the two-dimensional harmonic oscillator wave function with quantum numbers $n,m$. Inserting in~(\ref{J_osc}) the ground state wave function $X^{0,0}$ leads to the elastic scattering amplitude as given in~\cite{Berger}.

\begin{figure}[h]
  \unitlength0.8mm
  \begin{center}
    \begin{picture}(85,85)
      \put(0,0){\epsfysize=80mm \epsffile{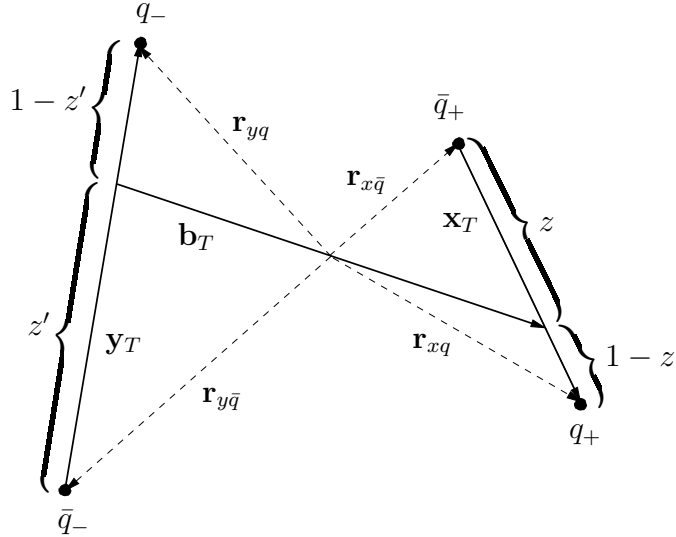}}
      \put(62.5,64){$\bar q_+$}
      \put(85,10){$q_+$}
      \put(0,-5){$\bar q_-$}
      \put(13,80){$q_-$}
      \put(88,15){
      \begin{rotate}{116}
        $\underbrace{\hspace{12mm}}\underbrace{\hspace{27mm}}$
      \end{rotate}}
      \put(80,44.5){$z$}
      \put(91,22){$1-z$}
      \put(64,45){${\bf x}_T$}
      \put(-3,1){
      \begin{rotate}{81}
        $\overbrace{\hspace{41.5mm}}\overbrace{\hspace{18.5mm}}$
      \end{rotate}}
      \put(-8,64.5){$1-z^\prime$}
      \put(-5,27){$z^\prime$}
      \put(8,25){${\bf y}_T$}
      \put(20,42){${\bf b}_T$}
      \put(29,61){${\bf r}_{yq}$}
      \put(24,16){${\bf r}_{y\bar q}$}
      \put(59,25){${\bf r}_{xq}$}
      \put(48,52){${\bf r}_{x\bar q}$}
    \end{picture}
  \end{center}
  \vspace*{-0.2cm}
  \caption{{\em Orientation and extension of the light-like Wegner-Wilson loops in a projection into transverse position space}}\label{vecs_fig}
\end{figure}

In~(\ref{J_wave}) and~(\ref{J_osc}) $w_{31}({\bf x}_T,z)$ denotes the profile function for the overlap between initial and final state of hadron $h_1$ for fixed ${\bf x}_T$ and $z$. $\varphi_2({\bf y}_T,z^\prime)$ defines the initial state wave function of $h_2$.
The light-like Wegner-Wilson loops $\mathcal{W}_\pm$ are given by
\begin{equation} \label{loop}
  \mathcal{W}_\pm:=\frac{1}{3} \,\mbox{tr}\, V(C_\pm) = \frac{1}{3} \,\mbox{tr P}\exp{(-ig \int_{C_\pm} \mbox{d}x^\mu G^a_\mu(x) \frac{\lambda^a}{2})},
\end{equation}
where P denotes path ordering and $C_\pm$ is the curve consisting of two light-like worldlines for the quark and the antiquark and connecting pieces at $\pm \infty$. In~(\ref{J_wave}),(\ref{J_osc}) ${\bf x}_T$ and ${\bf y}_T$ define the extension and orientation in transverse position space of the two loops representing the two hadrons $h_1$ and $h_2$ respectively, $z$ ($z^\prime$) parametrises the fraction of the longitudinal momentum of hadron $h_1$ ($h_2$) carried by the quark.
The impact parameter is given by ${\bf b}_T$, the light-cone barycentres of the loops are then located at $\frac{1}{2}{\bf b}_T+(\frac{1}{2}-z){\bf x}_T$ and $-\frac{1}{2}{\bf b}_T+(\frac{1}{2}-z^\prime){\bf y}_T$, respectively (see~\cite{DGKP} and Fig.~\ref{vecs_fig}). As $x$-axis for the transverse vectors ${\bf x}_T, \, {\bf y}_T$ and ${\bf b}_T$ we choose ${\bf q}_T$.\\
The symbol $\langle\ldots\rangle_G$ denotes the functional integration which correlates the two loops. In~(\ref{J_wave}) the loop-loop correlation function is multiplied with the profile function $w_{31}$, the incoming wave function $\varphi_2$ and then integrated over all extensions and orientations of the loops in transverse space as well as over the longitudinal momentum fraction $z$ of the quark of hadron $h_1$, which does not break up diffractively.
In the plane wave description $z^\prime$ is part of the specification of the final state and thus appears as arguement of $\hat J_{\mbox{\scriptsize{diff}}}^{\mbox{\scriptsize{p.wave}}}({\bf b}_T,z^\prime)$ in~(\ref{J_wave}). The phase space integral then includes an integration over $z^\prime$.
When using the second method involving the two-dimensional oscillator functions to describe the diffractive final state, one has to insert the function $X^{n,m}$ on the r.h.s. of~(\ref{J_osc}) and to integrate over $z^\prime$.
Thus $\hat J_{\mbox{\scriptsize{diff}}}^{\mbox{\scriptsize{2d osc.}}}({\bf b}_T)$ depends for given oscillator function $X^{n,m}$ on ${\bf b}_T$ only.

\subsection{Photo- and electroproduction}

For the reactions
\begin{eqnarray}
 \gamma \, + \, p &\to& \rho^0 \, + \, X, \nonumber \\
 \gamma^* \, + \, p &\to& \rho^0 \, + \, X,\label{gp_scat}
\end{eqnarray}
where $\gamma$ ($\gamma^*$) denotes a real (virtual) photon,
all equations derived in the previous section remain unchanged but of course
$w_{31}({\bf x}_T, z)$ in (\ref{J_wave}) now denotes the profile function for the overlap
between incoming photon and outgoing meson.
We use the wave functions for photon and $\rho^0$ meson as in \cite{DGKP}.
The reaction can again be depicted as shown in Fig.~\ref{hh_scat_pic},
with the incoming hadron $h_1(P_1)$ replaced by a (virtual) photon and
the outgoing hadron $h_1(P_3)$ by the $\rho^0$ meson.
The upper vertex now symbolises the dissociation of the photon into a quark-antiquark-pair
and the following formation of the vector meson out of these partons.
Again the momentum transfer is purely transverse and given by ${\bf q}_T$.
As before, $X$ denotes all possible diffractive final states into which the hadron $h_2$ can go.
Taking all this into account, the scattering amplitude~(\ref{scat_amp}),(\ref{J_wave}) applies to this type of reaction as well.


\bigskip
\newpage

\section{Evaluation of the scattering amplitudes}\label{eval_sect}

\subsection{The loop-loop correlation function in the model of the stochastic vacuum}

Now we perform the functional integrals in~(\ref{J_wave}) and~(\ref{J_osc}) making use of the MSV.\\
A detailed presentation of the MSV can be found in~\cite{DFK,MSV,MSVmink,DRio}, where both the original formulation in Euclidian space-time and the analytic continuation to Minkowski space-time are discussed. Here we will focus on some fundamental properties which are relevant for the evaluation of the expectation value of the loop-loop correlation function.\\
The starting point for the model is the correlator of two gluon field strength tensors $G_{\mu\nu}^a$ at points $x_1$ and $x_2$, parallel-transported to a common reference point $o$ along the two curves $C_{x_1}$ and $C_{x_2}$:
\begin{equation}
  \Bigl\langle \frac{g^2}{4 \pi^2} \hat G_{\mu \nu}^a(o,x_1;C_{x_1}) \hat G^{\mu \nu b}(o,x_2;C_{x_2}) \Bigr\rangle_G \equiv \frac{1}{4} \delta^{ab} F_{\mu\nu\rho\sigma}(x_1,x_2,o;C_{x_1},C_{x_2}) \label{ll_correl}
\end{equation}
The right hand side depends only on the points $x_1, \, x_2$ and the two curves $C_{x_1}, \, C_{x_2}$, the common reference point $o$ can be freely shifted along the curve $C_{12}=C_{x_1}+\bar C_{x_2}$, where $\bar C_{x_2}$ denotes the curve oriented oppositely to $C_{x_2}$.
Due to colour conservation, the correlation function is proportional to $\delta^{ab}$.
In the MSV the assumption is made that $F_{\mu\nu\rho\sigma}$ is independent of the choice of the connecting curve $C_{12}$. Then Poincar\'e and parity invariance require $F_{\mu\nu\rho\sigma}$ to be of the following form:
\begin{eqnarray}\label{correl_ansatz}
  F_{\mu\nu\rho\sigma}(z)&=&\frac{1}{24}G_2 \biggl\{ ( g_{\mu\rho}g_{\nu\sigma}-g_{\mu\sigma}g_{\nu\rho}) \left[ \kappa D(z^2)+(1-\kappa)D_1(z^2) \right] \nonumber \\
& &+(z_\sigma z_\nu g_{\mu\rho}-z_\rho z_\nu g_{\mu\sigma}+z_\rho z_\mu g_{\nu\sigma} - z_\sigma z_\mu g_{\nu\rho})(1-\kappa) \frac{\mbox{d}D_1(z^2)}{\mbox{d}z^2} \biggr\},\\
G_2 &\equiv& \frac{1}{4\pi^2} \, \langle g^2 FF \rangle = \langle 0 |  \frac{g^2}{4\pi^2} \, G_{\mu\nu}^a(0)  G^{a\mu\nu}(0) |0\rangle,
\end{eqnarray}
where $z=x_1-x_2$. Here $G_2$ is proportional to the gluon condensate $\langle 0 | G_{\mu\nu}^a(0) G^{a\mu\nu}(0) \, |0\rangle$, $D$ and $D_1$ are invariant functions normalised to 1 at $z=0$, $D(0)=D_1(0)=1$, and $\kappa$ is a parameter determining the non-abelian character of the correlator. The properties of the functions $D$ and $D_1$ are further specified through the assumption of the MSV that for space-like separations those functions rapidly fall to zero on a scale given by the correlation length $a \approx 0.3 \mbox{ fm}$. The Fourier decomposition of those functions is given by
\begin{eqnarray}
  D(z^2) &=& \int_{-\infty}^\infty \frac{\mbox{d}^4k}{(2 \pi)^4} e^{-ikz} \tilde D(k^2), \nonumber \\
  D_1(z^2) &=& \int_{-\infty}^\infty \frac{\mbox{d}^4k}{(2 \pi)^4} e^{-ikz} \tilde D_1(k^2). \label{D}
\end{eqnarray}
A suitable ansatz for $\tilde D$ and $\tilde D_1$ is given in~\cite{DFK}:
\begin{eqnarray}
  \tilde D(k^2) &=& \frac{27(2 \pi)^4}{(8a)^2} \frac{ik^2}{(k^2-\lambda^{-2}+i\epsilon)^4}, \nonumber \\
  \tilde D_1(k^2) &=& \frac{2}{3} \frac{27(2 \pi)^4}{(8a)^2} \frac{i}{(k^2-\lambda^{-2}+i\epsilon)^3},  \label{Dtilde}
\end{eqnarray}
with the constant $\lambda=8a/3\pi$. The functions of~(\ref{D}),(\ref{Dtilde}) can be compared to lattice calculations~\cite{correl_lattice,Meggiolaro} for the Euclidian version of the correlator~(\ref{ll_correl}) and from a fit one can extract the following ranges for the parameters $G_2,a,\kappa$~\cite{Meggiolaro}:
\begin{eqnarray} \label{msvpar}
  \kappa G_2 a^4 &=& 0.39 \, \dots \, 0.41, \nonumber \\
  \kappa &=& 0.80 \, \dots \, 0.89,         \nonumber \\
  a &=& 0.33 \, \dots \, 0.37 \mbox{ fm}.
\end{eqnarray}
The expectation value of one single parallel-transported gluon field strength tensor vanishes due to colour conservation and the fact that the QCD vacuum has no preferred direction in colour space:
\begin{equation}
  \langle \hat G(i) \rangle_G = 0,
\end{equation}
where we have used the abbrevation
\begin{equation}
  \hat G(i) \equiv \hat G^{a_i}_{\mu_i\nu_i}(o,x_i;C_{x_i}).
\end{equation}
Furthermore the MSV assumes the vacuum fluctuations of the field strengths to be determined by a Gaussian process. This implies that correlators of more than two gluon field strengths factorise and thus the process is completely defined by the second moment of its distribution.
Due to the assumption of a Gaussian process and colour conservation all $n$-point functions with odd $n$ vanish and we are left with
\begin{equation}
  \langle \hat G(1) \dots \hat G(2n) \rangle_G = \sum_{\mbox{\scriptsize all pairings}} \langle \hat G(i_1) \hat G(i_2) \rangle_G \dots \langle \hat G(i_{2n-1}) \hat G(i_{2n}) \rangle_G.
\end{equation}

\bigskip

\begin{figure}[h]
  \unitlength1mm
  \begin{center}
    \begin{picture}(90,95)
      \put(0,0){\epsfysize=90mm \epsffile{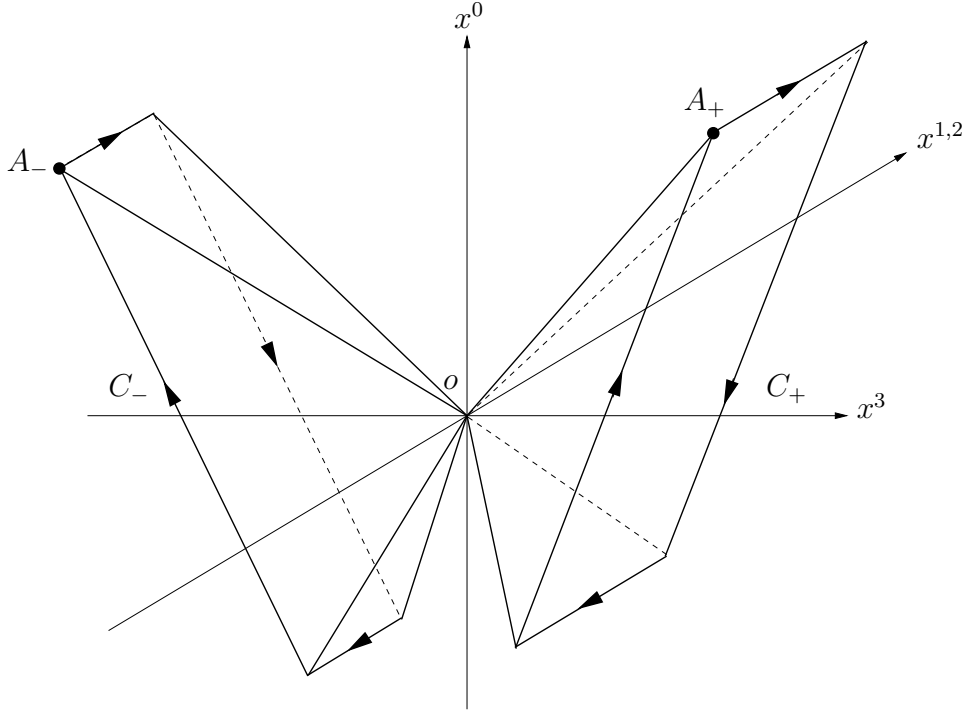}}
      \put(107,39){$x^3$}
      \put(53.5,91){$x^0$}
      \put(115,75){$x^{1,2}$}
      \put(95,42){$C_+$}
      \put(7.5,42){$C_-$}
      \put(-6,72){$A_-$}
      \put(84,80){$A_+$}
      \put(52,43){$o$}
    \end{picture}
  \end{center}
  \vspace*{-0.5cm}
  \caption{\em Integration surfaces for the evaluation of the loop-loop correlation function} \label{looploop_fig}
\end{figure}

Now we will make a cumulant expansion~\cite{Nachtmannreport} for the loop-loop correlation function 
\begin{eqnarray}
  \Bigl\langle \mathcal{W}_+ \mathcal{W}_- \Bigr\rangle_G \equiv \Bigl \langle \mathcal{W}_+(\frac{1}{2}{\bf b}_T+(\frac{1}{2}-z){\bf x}_T,{\bf x}_T) \mathcal{W}_-(-\frac{1}{2}{\bf b}_T+(\frac{1}{2}-z^\prime){\bf y}_T,{\bf y}_T)\Bigr \rangle_G\label{WW_eq}
\end{eqnarray}
in~(\ref{J_wave}),(\ref{J_osc}) and then evaluate our result in the framework of the MSV. \\
To expand the correlation function, we proceed as explained in~\cite{Berger}. First the line integrals along the closed loops $C_\pm$ are transformed to surface integrals with the help of the non-abelian Stokes theorem where, following the authors of~\cite{DFK}, we choose the mantle of a double pyramid as the integration surface. The basis surfaces $S_\pm$ of the two pyramids are enclosed by the two loops $C_\pm$ (see Fig.~\ref{looploop_fig}).
The common reference point $o$ is chosen to be the apex, where both pyramids touch, and $P_+$ and $P_-$ are the mantle surfaces of the two pyramids, respectively.
Following~\cite{Berger} we rewrite the two traces over $3 \times 3$ matrices occuring in~(\ref{WW_eq}) after inserting~(\ref{loop}) as one trace $(\mbox{Tr}_2)$ of a matrix acting in the 9-dimensional tensor product space.
With the definition
\begin{eqnarray} \label{total_G}
  \hat G_{t,\mu\nu}(o,x;C_x):=\left\{\begin{array}{lcl} \hat G_{\mu\nu}^a(o,x;C_x)(\frac{\lambda_a}{2} \otimes \mathbbm{1}) & \mathrm{for} & x \in P_+ \\ \hat G_{\mu\nu}^a(o,x;C_x)(\mathbbm{1} \otimes \frac{\lambda_a}{2}) & \mathrm{for} & x \in P_- \end{array} \right. ,
\end{eqnarray}
we can now write~(\ref{WW_eq}) as the expectation value of one ordered exponential in the product space, where the integration surface is given by the mantle $P=P_+ \cup P_-$ of the double pyramid:
\begin{eqnarray}
  \left\langle \mathcal{W}_+ \mathcal{W}_-\right\rangle_G=\frac{1}{9} \, \mbox{Tr}_2 \, \left\langle \mbox{P} \, \mbox{exp} \, \left(-\frac{ig}{2} \int_P \mbox{d}\sigma^{\mu\nu}(x) \, \hat G_{t,\mu\nu}(o,x;C_x) \right)\right\rangle_G.
\end{eqnarray}
The cumulant expansion of this expression up to the second term reads
\begin{eqnarray} \label{cumul}
  \lefteqn{\Bigl \langle \mathcal{W}_+(\frac{1}{2}{\bf b}_T+(\frac{1}{2}-z){\bf x}_T,{\bf x}_T) \mathcal{W}_-(-\frac{1}{2}{\bf b}_T+(\frac{1}{2}-z^\prime){\bf y}_T,{\bf y}_T)\Bigr \rangle_G} \nonumber\\
  & &= \frac{1}{9} \, \mbox{Tr}_2 \, \mbox{exp} \, \left(-\frac{g^2}{8}\int_P \mbox{d}\sigma^{\mu\nu}(x) \int_P \mbox{d}\sigma^{\rho\sigma}(x^\prime) \, \left\langle \mbox{P} ( \hat G_{t,\mu\nu}(o,x;C_x) \hat G_{t,\rho\sigma}(o,x^\prime;C_{x^\prime}) ) \right\rangle_G \right) \nonumber \\
  & &=:\frac{1}{9}\mathrm{Tr}_2 \, \mbox{exp} \; C_2({\bf b}_T,{\bf x}_T,{\bf y}_T,z,z^\prime),
\end{eqnarray}
where $C_2$ is a $9 \times 9$ matrix invariant unter $SU(3)$ colour rotations. As shown in~\cite{Berger} this finally leads to
\begin{eqnarray}
  \left\langle \mathcal{W}_+ \mathcal{W}_-\right\rangle_G=\frac{2}{3}e^{-i\frac{1}{3}\chi}+\frac{1}{3}e^{i\frac{2}{3}\chi}
\end{eqnarray}
with
\begin{eqnarray} \label{chi}
   \chi({\bf b}_T,{\bf x}_T,{\bf y}_T,z,z^\prime) &=& \frac{G_2 \pi^2}{24} \left\{ I({\bf r}_{xq},{\bf r}_{yq}) + I({\bf r}_{x \bar q},{\bf r}_{y \bar q}) - I({\bf r}_{xq},{\bf r}_{y \bar q}) - I({\bf r}_{x \bar q},{\bf r}_{yq}) \right\}, \nonumber \\
   I({\bf r}_x,{\bf r}_y) &=& \kappa \frac{\pi}{2} \lambda^2 \; {\bf r}_y \cdot {\bf r}_x \int_0^1 \mbox{d}v \, \Biggl\{ \left(\frac{|v{\bf r}_y-{\bf r}_x|}{\lambda}\right)^2 K_2\left(\frac{|v{\bf r}_y-{\bf r}_x|}{\lambda}\right) \nonumber \\
& & \hspace{3cm}+\left(\frac{|{\bf r}_y-v{\bf r}_x|}{\lambda}\right)^2 K_2\left(\frac{|{\bf r}_y-v{\bf r}_x|}{\lambda}\right)\Biggr\} \nonumber\\
& &+(1-\kappa)\pi\lambda^4 \left(\frac{|{\bf r}_y-{\bf r}_x|}{\lambda}\right)^3 K_3\left(\frac{|{\bf r}_y-{\bf r}_x|}{\lambda}\right).
\end{eqnarray}
Here $G_2, \lambda, \kappa$ are as defined in~(\ref{correl_ansatz}),(\ref{Dtilde}) and $K_{2,3}$ are the modified Bessel functions of second and third degree. The vectors ${\bf r}_{ij}$ with $i=x,y$ and $j=q,\bar q$ are those from the coordinate origin to the positions of the quarks and antiquarks (or diquarks) in transverse space as shown in Fig.~\ref{vecs_fig}.
Separating the real and the imaginary part of the above expression ($\chi$ is a real function) gives
\begin{eqnarray}\label{eikonal_result}
   \lefteqn{\left\langle \mathcal{W}_+(\frac{1}{2}{\bf b}_T+(\frac{1}{2}-z){\bf x}_T,{\bf x}_T) \mathcal{W}_-(-\frac{1}{2}{\bf b}_T+(\frac{1}{2}-z^\prime){\bf y}_T,{\bf y}_T) - \mathbbm{1} \right\rangle_G} \hspace{1cm} \nonumber \\
&=& \; \Biggl\{ \frac{2}{3}\cos\left(\frac{1}{3}\chi({\bf b}_T,{\bf x}_T,{\bf y}_T,z,z^\prime)\right)+\frac{1}{3}\cos\left(\frac{2}{3}\chi({\bf b}_T,{\bf x}_T,{\bf y}_T,z,z^\prime)\right) -1 \nonumber \\
& &-i\frac{2}{3}\sin\left(\frac{1}{3}\chi({\bf b}_T,{\bf x}_T,{\bf y}_T,z,z^\prime)\right)+i\frac{1}{3}\sin\left(\frac{2}{3}\chi({\bf b}_T,{\bf x}_T,{\bf y}_T,z,z^\prime)\right) \Biggr\}.
\end{eqnarray}
This is the final result for the correlation function of two light-like Wegner-Wilson loops in the matrix cumulant method~\cite{Berger}. If we assume $|\chi| \ll 1$,~(\ref{eikonal_result}) reduces to 
\begin{equation}\label{chi2_result}
  \left\langle \mathcal{W}_+ \mathcal{W}_- - \mathbbm{1}\right\rangle_G=\left\{ -\frac{1}{9} \left(\chi({\bf b}_T,{\bf x}_T,{\bf y}_T,z,z^\prime)\right)^2 \right\},
\end{equation}
neglecting terms of order $\chi^3$ and higher. This is the result of the traditional expansion method~\cite{DFK}. When computing the numerical results for the cross sections we are interested in, we will use both~(\ref{eikonal_result}) and (\ref{chi2_result}) and compare with experimental data. In the following formulae, we will write $\{ \ldots \}$ as an abbrevation for either of the expressions~(\ref{eikonal_result}) or~(\ref{chi2_result}).

\subsection{Hadron-hadron scattering}\label{eval_hh_sect}
We now have to specify the hadronic wave functions occuring in~(\ref{J_wave}),(\ref{J_osc}).
For the hadronic wave functions we make the following ansatz:
\begin{eqnarray}
  \varphi_i({\bf x}_T,z) &=& \sqrt{\frac{2 z (1-z)}{2 \pi S_{h_i}^2 I_{h_i}}} \,  e^{-(z-\frac{1}{2})^2/4 z_{h_i}^2} \, e^{-{\bf x}_T^2/4 S_{h_i}^2}, \label{wave_funct} \\
  w_{ij}({\bf x}_T,z) &=& \varphi_i({\bf x}_T,z) \, \varphi_j({\bf x}_T,z).
\end{eqnarray}
Here $I_{h_i}$ is a normalisation factor given by
\begin{equation}\label{norm_z}
  I_{h_i}=\int_0^1 \mbox{d}z \, 2 z (1-z) \, e^{-(z-\frac{1}{2})^2/2 z_{h_i}^2}.
\end{equation}
The wave functions $X^{n,m}$ consist of the eigenfunctions $\tilde X^{n,m}$ of a two-dimensional harmonic oscillator~\cite{Cohen} for the ${\bf y}_T$-dependence and an additional part for the $z^\prime$-dependence as in~(\ref{wave_funct}):
\begin{eqnarray} \label{osc_funct}
  X^{n,m}({\bf y}_T,z^\prime) &=& \sqrt{\frac{2 z^\prime (1-z^\prime)}{I_{h_i}}} \, e^{-(z^\prime-\frac{1}{2})^2/4 z_{h_i}^2} \, \tilde X^{n,m}({\bf y}_T), \nonumber \\
  \tilde X^{n,m}({\bf y}_T) &=& \frac{1}{\sqrt{((n+m)/2) ! \, ((n-m)/2) !}} \, \left[\sqrt{\frac{S_{h_i}^2}{2}} \left( \frac{y_T}{2 S_{h_i}^2} + \frac{m}{y_T} -\frac{\mathrm{d}}{\mathrm{d} y_T} \right) \right]^{\frac{n+m}{2}} \nonumber \\ 
  & & \quad \left[\sqrt{\frac{S_{h_i}^2}{2}} \left( \frac{y_T}{2 S_{h_i}^2} - \frac{m}{y_T} - \frac{\mathrm{d}}{\mathrm{d} y_T} \right)\right]^{\frac{n-m}{2}} \frac{e^{-\beta^2 y_T^2 / 2}}{\sqrt{2 \pi S_{h_i}^2}} \, e^{i m \theta_y},
\end{eqnarray}
where $\theta_y$ is the angle between ${\bf y}_T$ and ${\bf q}_T$.
Now we put everything together, inserting the wave functions~(\ref{wave_funct})-(\ref{osc_funct}) and the results~(\ref{eikonal_result}),(\ref{chi2_result}) for the correlation function of the Wegner-Wilson loops in~(\ref{J_wave}) and~(\ref{J_osc}). We can simplify our expressions by exploiting symmetry properties of the wave and correlation funtions.
The replacements ${\bf x}_T \to -{\bf x}_T$ and $z \to 1-z$, which exchange the quark with the corresponding diquark (or antiquark in the case of mesons) in hadron $h_1$, lead to $\chi \to -\chi$ (see Fig.~\ref{vecs_fig} and~(\ref{chi})). On the other hand these replacements leave the wave functions invariant and thus the integration over ${\bf x}_T$ and $z$ averages out the $\sin \chi$-terms of~(\ref{eikonal_result}) when inserted in~(\ref{J_wave}) and~(\ref{J_osc}).
For single diffractive dissociation and elastic scattering we can therefore replace~(\ref{eikonal_result}) by
\begin{equation}\label{eikonal_2}
   \left\langle \mathcal{W}_+\mathcal{W}_- - \mathbbm{1} \right\rangle_G \to  \left \{ \frac{2}{3}\cos \left(\frac{1}{3}\chi \right)+\frac{1}{3}\cos \left(\frac{2}{3}\chi \right) -1\right \}.
\end{equation}
In the expansion method $\left\langle \mathcal{W}_+\mathcal{W}_- - \mathbbm{1} \right\rangle_G$ in~(\ref{chi2_result}) is already even under $\chi \to - \chi$.
In our model, therefore, the expression for the correlation function is purely real in~(\ref{chi2_result}) and only the real part of~(\ref{eikonal_result}) contributes. The $\mathcal{T}$-matrix element is invariant under the exchange of hadron $h_1$ by its antihadron. Thus we get only $C=+1$ (pomeron) exchange and no $C=-1$ (odderon) exchange.
An imaginary part of the correlation function and $C=-1$ exchange terms both non-vanishing after integration with the overlap functions could arise from the inclusion of higher cumulant terms in~(\ref{cumul}). We note that in double diffractive reactions a $C=-1$ part of the scattering amplitude will already be present in our approximation with terms up to the second cumulant in the matrix cumulant method.

For the description of the final state $X$ in the oscillator method we can use analogous arguements.
The simultaneous replacements ${\bf y}_T \to -{\bf y}_T$ and $z^\prime \to 1-z^\prime$ and subsequent integration over ${\bf y}_T$ and $z^\prime$ lead to the cancellation of contributions with odd $m$ in~(\ref{J_osc}) because of the existence of a factor $e^{im\theta_y}$ in $\tilde X^{n,m}({\bf y}_T)$. Since for these functions odd $m$ only occur for odd $n$, the sum over all excited states in the calculation of cross sections can be reduced to a sum over wave functions with even $n$ and the corresponding $m$'s.\\

\subsection{Photo- and electroproduction of $\rho^0$ mesons}
%
As mentioned in chapter~\ref{scat_sect}, the main difference in our ansatz between the 
$\mathcal{T}$-matrix element of hadron-hadron scattering and the photo- and electroproduction of a vector meson
is in the wave functions used to describe the participating hadrons and the photon.
In electroproduction we have to consider three polarisation states for both the incoming virtual photon and the outgoing vector meson.
Since in our model processes with a change in polarisation are strongly suppressed (see \cite{DGKP}),
they will be neglected in the following.
So we only have to deal with photon and meson wave functions with the same polarisation.
Furthermore the expressions for the two transverse cases are the same, which leaves us with just two combinations:
\begin{eqnarray}
  w_{31,\mathrm{L}}({\bf x}_T, z)  
   &=&  \frac{1}{4\pi}\sum_{h,h'} \left( \varphi^{\,\rho^0}_{h,h',V(0)}(z,{\bf x}_T) \right)^*
                                  \varphi^{\gamma^{(*)}}_{h,h'\,(Q^2 , 0)}(z , {\bf x}_T)                           \label{wf_rpL} \\ 
   &=&    -\frac{1}{4\pi}ef_\rho z(1-z) f_\mathrm{L}(z) e^{-\omega^2_\mathrm{L} x_T^2 / 2 } 2 z(1-z) Q K_0( \epsilon x_T ) \nonumber \;, \\[2ex]
  w_{31,\mathrm{T}}({\bf x}_T, z)  
   &=&  \frac{1}{4\pi}\sum_{h,h'} \left( \varphi^{\,\rho^0}_{h,h',V(1/-1)}(z,{\bf x}_T) \right)^*
                                  \varphi^{\gamma^{(*)}}_{h,h'\,(Q^2 ,1/-1)}(z ,{\bf x}_T)                          \label{wf_rpT}  \\ 
   &=& \!\!\!\! -\frac{1}{4\pi}ef_\rho f_\mathrm{T}(z) e^{-\omega^2_\mathrm{T} x_T^2 / 2 }                                    \nonumber      \\
   &&      \qquad   \times   \left\{ \frac{\omega^2_\mathrm{T} \epsilon x_T}{M_\rho} \left[ z^2 + (1-z)^2 \right] K_1(\epsilon x_T) 
                      +\frac{m(Q^2)^2}{M_\rho} K_0(\epsilon x_T) \right\} \nonumber \;,
\end{eqnarray}
with
\begin{eqnarray}
  \epsilon &=& \sqrt{ z(1-z)Q^2 + m(Q^2)^2 \,}  \; , \\[2ex]
  f_\mathrm{L/T}(z) 
   &=& {\cal N}_\mathrm{L/T} \sqrt{z(1-z)} \, \exp\left( -\frac{M^2_\rho}{2\omega^2_\mathrm{L/T}}\left(z-\frac{1}{2}\right)^2 \right) \; .
\end{eqnarray}
Here $Q^2$ denotes the virtuality of the photon.
Further parameters are the proton charge $e$,
the mass of the $\rho^0$ meson $M_\rho = 768.5\mbox{ MeV}$,
its decay constant $f_\rho = 150.7 \mbox{ MeV}$
and its radius parameters and normalisation constants $\omega_\mathrm{L/T}$ and ${\cal N}_\mathrm{L/T}$, which are different for the longitudinal and
the transverse cases. For the values of these parameters and how to fix them see~\cite{DGKP}.\\
\\
In model investigations it has been shown that confinement effects can be very well approximated
and the photon densities can be extended to real photons
by a $Q^2$-dependent light quark mass \cite{DGP98}
%
\begin{eqnarray}
  m(Q^2) &=& \left\{ \begin{array}{rl} \left(1-\frac{Q^2}{Q_0^2}\right)0.22\mbox{ GeV}  & \mbox{for } Q^2  < Q_0^2      \\
                   0                                                                    & \mbox{for } Q^2 \ge Q_0^2    \end{array}
               \right.
  \quad,\quad  Q_0^2 = 1.05\mbox{ GeV}^2 \;.
\end{eqnarray}
Of course for photoproduction, that is for $Q^2=0$, the longitudinal overlap function~(\ref{wf_rpL}) vanishes, as it must be since real photons have only transverse polarisation.

%


\bigskip
\section{Total and differential cross sections}

To calculate cross sections for the reactions considered, we have to fix our free parameters, namely those of the MSV: $G_2$, $a$ and $\kappa$; and those of the wave functions, the extension parameter $S_{h_i}$ and the width of the longitudinal momentum distribution $z_{h_i}$. The set of MSV parameters used in this work has been established in~\cite{Berger} for the case of the matrix cumulant method giving~(\ref{eikonal_result}) and in~\cite{DGKP} for the expansion method giving~(\ref{chi2_result}) (see Table~\ref{par_tab}).
\begin{center}
\begin{table}[H]
\begin{tabular}{|c|c|c|}
  \hline
  & matrix method & expansion method \\ \hline
  $G_2$ & $(529 \mbox{ MeV})^4$ & $(501 \mbox{ MeV})^4$ \\
  $a$   & $0.32 \mbox{ fm}$     & $0.346 \mbox{ fm}$ \\
  $\kappa$ & $0.74$             & $0.74$ \\ \hline
\end{tabular}
\caption{\em The parameters of the MSV for the matrix cumulant and the expansion methods}\label{par_tab}
\end{table}
\end{center}
The values given in Table~\ref{par_tab} should be considered as effective values extracted from fits to high energy scattering data using two different approximate formulae. Thus the differences between the values in the second and third column of the table can be taken as a theoretical error estimate.
With fixed parameters the model gives energy independent cross sections. It has been shown in~\cite{DFK} that both the energy dependence of the cross section and of the slope parameter $b$ of elastic scattering can be well described by energy dependent hadron extension parameters $S_{h_i}(s)$. In~\cite{Berger} it was found that in the framework of the matrix cumulant method energy dependent extension parameters can even describe the energy evolution of the whole differential elastic cross sections $\mathrm{d}\sigma / \mathrm{d} t$ up to $|t| \approx 1 \, \mathrm{GeV}^2$. For the matrix cumulant method we adopt the parametrisation from~\cite{Berger}:
\begin{equation}\label{S2_eikonal}
  S_p(s) = 0.700 \left( \frac{s}{\mbox{GeV}^2} \right)^{0.034} \mbox{ fm}.
\end{equation}
This was obtained by fitting the total cross section as calculated in the model to the soft pomeron part of the Donnachie-Landshoff (DL) parametrisation for $\sigma_{\mbox{\scriptsize{tot}}}$~\cite{DL}. For the expansion method we have established a similar connection between $S_p$ and $s$:
\begin{equation}\label{S2_chi2}
  S_p(s) = 0.624 \left( \frac{s}{\mbox{GeV}^2} \right)^{0.028} \mbox{ fm}.
\end{equation}
At $\sqrt{s}=23.5 \mbox{ GeV}$, for instance, we get $S_p=0.868 \mbox{ fm}$ and $S_p=0.745 \mbox{ fm}$ from~(\ref{S2_eikonal}) and (\ref{S2_chi2}), respectively.
The width of the longitudinal momentum distribution of the proton has been chosen as $z_p=0.4$ which gives a best fit to the isovector form factor of the proton calculated in the framework of our model (see chapter~\ref{ff_sect}).

\subsection{Proton-proton scattering}\label{results_pp_sect}

\subsubsection{Elastic scattering}
\begin{figure}[h]
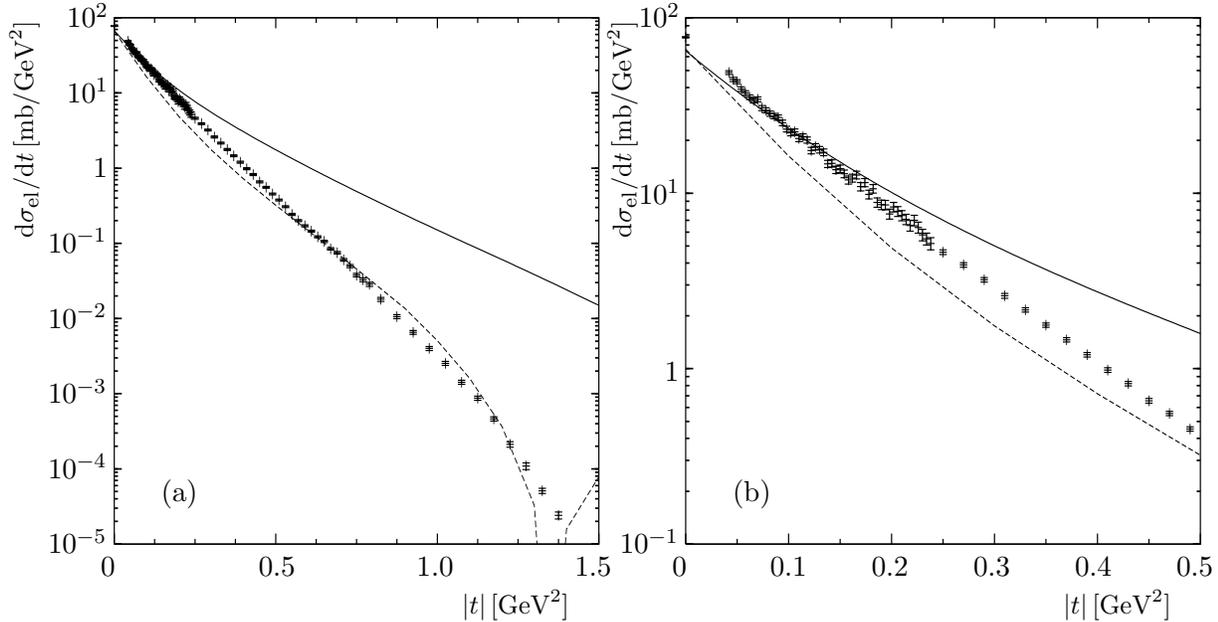

  \unitlength1mm
  \begin{center}
    \begin{picture}(160,80)
      \put(0,0){\epsfig{file=fig.3a,height=80mm,width=80mm}}
      \small
      \put(10,2){$0$}
      \put(31,2){$0.5$}
      \put(52,2){$1.0$}
      \put(73,2){$1.5$}
      \put(3,5){$10^{-5}$}
      \put(3,15){$10^{-4}$}
      \put(3,25){$10^{-3}$}
      \put(3,35){$10^{-2}$}
      \put(3,45){$10^{-1}$}
      \put(9,55){$1$}
      \put(5.5,65){$10^1$}
      \put(5.5,75){$10^2$}
      \put(18,12){(a)}
      \footnotesize
      \put(58,-3){$|t| \, [\mbox{GeV}^2]$}
      \put(1,47){\begin{rotate}{90} $\mbox{d}\sigma_{\mbox{\scriptsize{el}}} / \mbox{d}t \, [\mbox{mb}/\mbox{GeV}^2]$ \end{rotate}}
      \put(80,0){\epsfig{file=fig.3b,height=80mm,width=80mm}}
      \small
      \put(86,2){$0$}
      \put(99,2){$0.1$}
      \put(112.5,2){$0.2$}
      \put(126,2){$0.3$}
      \put(139.5,2){$0.4$}
      \put(153,2){$0.5$}
      \put(79,5){$10^{-1}$}
      \put(85,28){$1$}
      \put(81.5,51.5){$10^1$}
      \put(81.5,75){$10^2$}
      \put(94,12){(b)}
      \footnotesize
      \put(138,-3){$|t| \, [\mbox{GeV}^2]$}
      \put(80.5,47){\begin{rotate}{90} $\mbox{d}\sigma_{\mbox{\scriptsize{el}}} / \mbox{d}t \, [\mbox{mb}/\mbox{GeV}^2]$ \end{rotate}}
    \end{picture}
  \end{center}
  \vspace*{-0.5cm}
  \caption{{\em The differential elastic cross section} $\mbox{d}\sigma_{\mbox{\scriptsize{el}}} / \mbox{d}t \, [\mbox{mb}/\mbox{GeV}^2]$ {\em at} $\sqrt{s}=23.5 \mbox{ GeV}$ {\em calculated using the matrix cumulant method (dashed line) and the expansion method (solid line) compared to the experimental data from}~\cite{pp_elast_s23}}\label{dsigdt_fig}
\end{figure}
With all parameters fixed, we can now perform the numerical calculations for the scattering amplitude~(\ref{scat_amp})-(\ref{J_osc}). We will start with a short reminder of the results for elastic scattering obtained in~\cite{Berger} as these will be needed further on. As mentioned above, elastic scattering can be studied with our formulae when replacing the excited state wave function $X^{n,m}$ in~(\ref{J_osc}) by the ground state wave function $X^{0,0}$, which is identical to~(\ref{wave_funct}).
For $s \gg M_p^2$ the differential cross section is given by
\begin{equation} \label{dsig}
  \mbox{d}\sigma_{\mbox{\scriptsize{el}}}=\frac{1}{16 \pi} \frac{1}{s^2} \left| \mathcal{T}_{fi} \right|^2 \mbox{d}t.
\end{equation}
In~\cite{DGKP} it has been argued that the Gaussian shaped distribution of the longitudinal momentum fraction $z$ ($z^\prime$) can be replaced by a delta-function centred at $z=1/2$ ($z^\prime=1/2$), since the function $\chi$~(\ref{chi}), which determines the shape of the correlation function, depends only weakly on $z$ ($z^\prime$). A numerical investigation of the total cross section calculated from the optical theorem shows that the resulting difference for $\sigma_{\mbox{\scriptsize{tot}}}$ is smaller than 1\%. The profit one draws from this simplification is a much shorter computation time in the numerical analysis\footnote{All occuring integrals in the expressions for the cross sections are calculated using the Monte-Carlo integration subroutine {\sc Vegas}~\cite{Vegas}. As a rule of thumb, one has to use at least 10 datapoints per integration variable. The substitution thus decreases the computation time by roughly a factor of 100.}. In the following we will make use of this simplification if not explicitly stated otherwise.\\
In Fig.~\ref{dsigdt_fig} we compare the results from the matrix cumulant and expansion methods to experiment. The first method, i.e. using~(\ref{eikonal_2}), gives a reasonable description of the data for $|t| \lesssim 1 \, \mathrm{GeV}^2$ over many orders of magnitude but underestimates the data at small $|t|$. The expansion method, i.e. using~(\ref{chi2_result}), gives a better description of the data for $|t| \lesssim 0.2 \, \mathrm{GeV}^2$ but overshoots the data by orders of magnitude for larger $|t|$.
A fit of the form $\mathrm{d} \sigma_{\mbox{\scriptsize{el}}} / \mathrm{d} t = A \exp b \, t$ to the differential cross section gives $b= 13.8 \pm 0.4 \, \mathrm{GeV}^{-2}$ for the matrix cumulant method and $b= 10.0 \pm 0.2 \, \mathrm{GeV}^{-2}$ for the expansion method, respectively. From a fit to the experimental data~\cite{pp_elast_s23} we obtain $b= 11.6 \pm 0.1 \, \mathrm{GeV}^{-2}$.
These fits have been performed within the range $0 \le |t| \le 0.2 \, \mathrm{GeV}^2$, since the description of the data over a larger $|t|$-range would require an additional term $\propto t^2$ in the exponent of the fit.
If we calculate the integrated elastic cross section at $\sqrt{s}=23.5 \, \mathrm{GeV}$ we obtain $\sigma_{\mbox{\scriptsize{el}}}=5.0 \, \mathrm{mb}$ in the matrix cumulant method and $\sigma_{\mbox{\scriptsize{el}}}=7.3 \, \mathrm{mb}$ in the expansion method compared to an experimental value of $\sigma_{\mbox{\scriptsize{el}}}=6.81 \pm 0.19 \, \mathrm{mb}$~\cite{PDG}.
\begin{figure}[h]
  \unitlength1mm
  \begin{center}
    \begin{picture}(120,95)
      \put(0,0){\epsfysize=90mm \epsffile{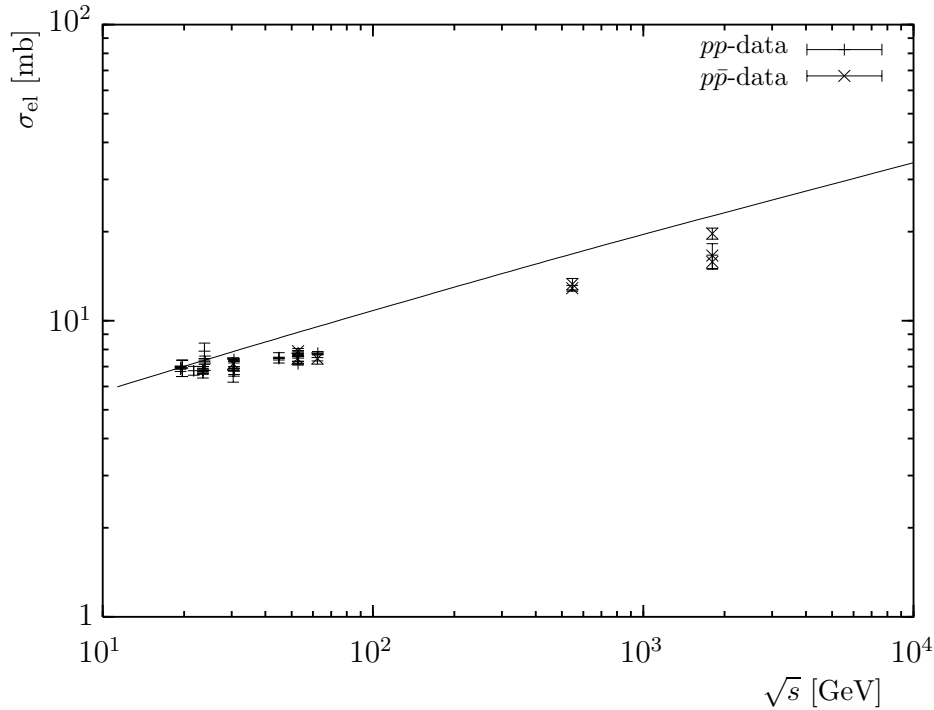}}
      \put(9,1){$10^1$}
      \put(45,1){$10^2$}
      \put(81,1){$10^3$}
      \put(117,1){$10^4$}
      \put(9,6){$1$}
      \put(5,45){$10^1$}
      \put(5,85){$10^2$}
      \small
      \put(100,-4){$\sqrt{s}$ [GeV]}
      \put(3,72){\begin{rotate}{90} $\sigma_{\mbox{\scriptsize{el}}}$ [mb]\end{rotate}}
      \footnotesize
      \put(90,82){$\begin{array}{r} \\ pp\mbox{-data} \\ p \bar p\mbox{-data} \end{array}$}
    \end{picture}
  \end{center}
  \vspace*{-0.3cm}
  \caption{{\em The integrated elastic cross section as a function of $\sqrt{s}$} {\em compared to the experimental data from}~\cite{PDG}}\label{sigel_fig}
\end{figure}
The fact that the elastic cross section calculated by the expansion method is closer to the experimental value than the one from the matrix cumulant method is easily understood from Fig.~\ref{dsigdt_fig}b. In the integral over $\mathrm{d}\sigma / \mathrm{d}t$ only the region $|t| \lesssim 0.2 \, \mathrm{GeV}^2$ contributes significantly and there the expansion method describes the data better.
In the region $|t| \gtrsim 0.2 \, \mathrm{GeV}^2$ the result from the expansion method is bigger than the experimental result, which has as a consequence that the resulting integrated cross section is too big.
In the following we will restrict our considerations to $\sigma_{\mbox{\scriptsize{tot}}}, \, \sigma_{\mbox{\scriptsize{el}}}$ and differential cross sections for $|t| \lesssim 0.2 \, \mathrm{GeV}^2$. For these quantities the expansion method gives a satisfactory description of the experimental data and we will use this method throughout the rest of our paper.
In Fig.~\ref{sigel_fig} we show $\sigma_{\mbox{\scriptsize{el}}}$ calculated in this way for $10 \mbox{ GeV} \le \sqrt{s} \le 10 \mbox{ TeV}$. The calculation agrees reasonably well with the experimental data of~\cite{PDG}. The data are as well from $pp$- as from $p \bar p$-experiments. Because our approach does not include, in Regge terminology, any non-leading trajectories, we cannot distinguish between these two reactions and they are described by the same scattering amplitude.

\subsubsection{Single diffractive dissociation}
\begin{figure}[h]
  \unitlength1mm
  \begin{center}
    \begin{picture}(120,90)
      \put(0,0){\epsfysize=90mm \epsffile{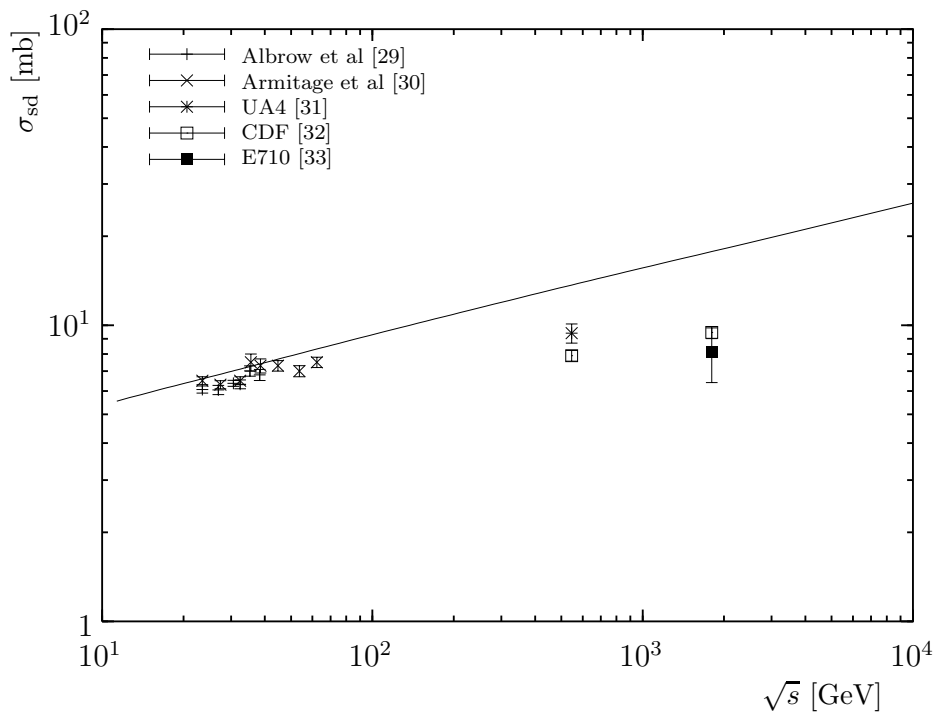}}
      \put(9,1){$10^1$}
      \put(45,1){$10^2$}
      \put(81,1){$10^3$}
      \put(117,1){$10^4$}
      \put(9,5){$1$}
      \put(5,45){$10^1$}
      \put(5,85){$10^2$}
      \small
      \put(100,-4){$\sqrt{s}$ [GeV]}
      \put(3,72){\begin{rotate}{90} $\sigma_{\mbox{\scriptsize{sd}}}$ [mb]\end{rotate}}
      \scriptsize
      \put(29,74.5){$\begin{array}{l} \mbox{Albrow et al}~\cite{Albrow} \\ \mbox{Armitage et al}~\cite{Armitage} \\ \mbox{UA4}~\cite{Bernard} \\ \mbox{CDF}~\cite{Abe} \\ \mbox{E710}~\cite{Amos} \end{array}$}
    \end{picture}
  \end{center}
  \vspace*{-0.3cm}
  \caption{{\em The integrated single diffractive dissociation cross section as a function of $\sqrt{s}$}}\label{sigdiss_fig}
\end{figure}

\begin{figure}[h]
  \unitlength1mm
  \begin{center}
    \begin{picture}(100,80)
      \put(0,0){\epsfig{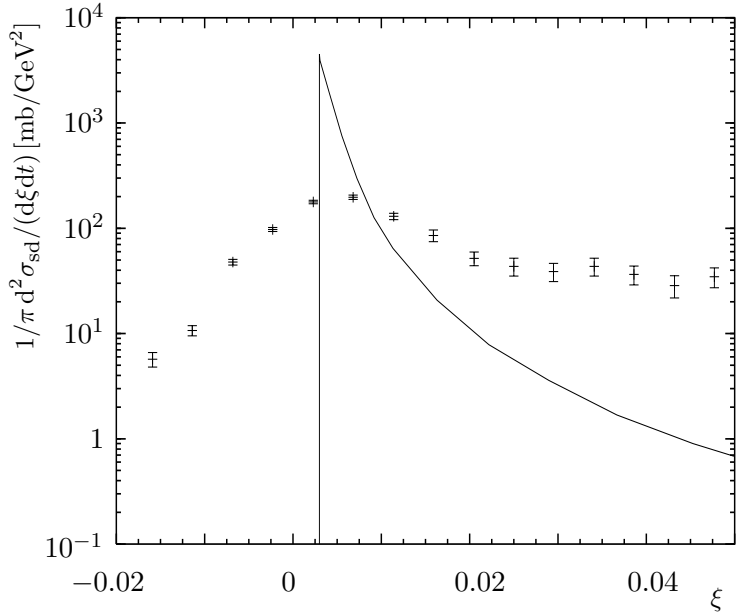}}
      \small
      \put(7,0){$-0.02$}
      \put(35,0){$0$}
      \put(58,0){$0.02$}
      \put(81,0){$0.04$}
      \put(3.5,4.5){$10^{-1}$}
      \put(9.5,19){$1$}
      \put(6,33){$10^1$}
      \put(6,47){$10^2$}
      \put(6,61){$10^3$}
      \put(6,75){$10^4$}
      \footnotesize
      \put(92,-2){$\xi$}
      \put(2,32){\begin{rotate}{90} $1 / \pi \, \mbox{d}^2\sigma_{\mbox{\scriptsize{sd}}} / (\mbox{d}\xi \mbox{d}t) \, [\mbox{mb}/\mbox{GeV}^2]$ \end{rotate}}
    \end{picture}
  \end{center}
  \vspace*{-0.5cm}
  \caption{{\em The differential diffractive cross section} $\mbox{d}^2\sigma_{\mbox{\scriptsize{sd}}} / (\mbox{d} \xi \mbox{d}t)$ {\em for } $t=-0.0525 \mbox{ GeV}^2$ {\em at } $\sqrt{s}=23.5 \mbox{ GeV}$ {\em together with data from}~\cite{Armitage}}\label{d2sig_fig}
\end{figure}

We now turn to inelastic diffractive scattering $p \: p \to p \: X$. Using the plane wave method (\ref{scat_amp}),(\ref{J_wave}), we calculate the differential diffractive cross section as
\begin{equation}
  \mbox{d}\sigma_{\mbox{\scriptsize{diff}}}=(2\pi)^4\frac{1}{2s} \left|\mathcal{T}_{fi}\right|^2 \mbox{d}^5\mathcal{P},
\end{equation}
where
\begin{eqnarray} \label{P5}
  \mbox{d}^5\mathcal{P}&=&\frac{1}{(2\pi)^9} \frac{1}{4s \, z^\prime (1-z^\prime)} \mbox{d}^2P_{4T} \mbox{d}^2\Delta_{4T} \mbox{d}z^\prime
\end{eqnarray}
is the 5-dimensional phase space measure for the three particle final state formed by the first proton which remains intact and the quark and the diquark which describe the second, diffractivly excited proton.
As stated above the description of the diffractive final state $X$ by a free quark-diquark pair also includes the case of elastic scattering. To obtain the cross section $\sigma_{\mathrm{sd}}$ for single diffractive dissociation, we have to subtract the elastic contribution and then multiply by 2 to account for the reaction where the first proton breaks up and the second stays intact. We then find for the integrated single diffractive cross section as a function of $\sqrt{s}$ the result shown in Fig.~\ref{sigdiss_fig}.
Comparing our results to experimental data, one has to keep in mind that the overall normalisation uncertainty of the experiments is of $\mathcal{O}(10\%)$. Furthermore the derivation of integrated cross sections from experimental data involves extrapolations of the measured data at given values of $t$ and $\xi=M_X^2/s$ to regions where no data exist. The extrapolations depend on assumptions on the shape of the $t$-distribution and the shape of the $\xi$-distribution. Different experiments make different assumptions and thus the resulting integrated cross sections differ from each other. The experimental values on the integrated single diffractive dissociation cross section quoted here use $\xi \le 0.05$ as an upper bound in the mass distribution~\cite{Albrow,Armitage,Bernard,Amos}, except for~\cite{Abe} where the range is extended to $\xi \le 0.2$. In our calculation of $\sigma_{\mathrm{sd}}$ we integrate over all values of $\xi$. Because the mass spectrum obtained in our calculation decreases rapidly with increasing $\xi$ (see Fig.~\ref{d2sig_fig}), our numerical result of the integrated cross section is dominated by the low mass region and is not sensible to the integration range being $\xi \le 0.05$ or $\xi \le 0.2$.

In Table~\ref{R_tab} we give the ratio $R$ of the single diffractive dissociation cross section to the sum of the single diffractive dissociation and the elastic cross sections from our model and from different experiments. For $\sqrt{s}=546 \, \mathrm{GeV}$ and 1800 GeV we have used the values of $\sigma_{\mathrm{el}}$ and $\sigma_{\mathrm{sd}}$ as quoted by the UA4, CDF and E710 experiments. For the ISR energy range $20 \, \mathrm{GeV} \lesssim \sqrt{s} \lesssim 60 \, \mathrm{GeV}$ a lot of data exist. Since the cross sections do not vary much over this energy range, we have fitted both $\sigma_{\mathrm{el}}$ and $\sigma_{\mathrm{sd}}$ as being proportional to a small power of $\sqrt{s}$ and have then calculated $R$ as a function of $\sqrt{s}$ using these fits. The quoted ISR $R$-value in Table~\ref{R_tab} is then evaluated at an intermediate energy of $\sqrt{s}=38.5 \, \mathrm{GeV}$.
\begin{table}[H]
\begin{center}
\begin{tabular}{|c|c|rclll|}
  \hline
  \multicolumn{7}{|c|}{$R=\sigma_{\mbox{\scriptsize{sd}}} / (\sigma_{\mbox{\scriptsize{el}}}+\sigma_{\mbox{\scriptsize{sd}}})$} \\ \hline
  $\sqrt{s}$ GeV & our calc. & \multicolumn{5}{c|}{values calc. from exp.} \\ \hline
  23.5 & 0.47 &      &       &      &     &                       \\
  38.5 & 0.47 & 0.49 & $\pm$ & 0.07 & ISR &\cite{Albrow,Armitage} \\
  62.3 & 0.46 &      &       &      &     &                       \\ \hline
  546  & 0.45 & 0.41 & $\pm$ & 0.02 & UA4 &\cite{Bernard} \\
       &      & 0.38 & $\pm$ & 0.01 & CDF &\cite{Abe} \\ \hline
  1800 & 0.44 & 0.33 & $\pm$ & 0.05 & E710&\cite{Amos}\\ 
       &      & 0.32 & $\pm$ & 0.01 & CDF &\cite{Abe} \\ \hline
\end{tabular}
\caption{\em The ratio $R$ of the single diffractive dissociation to the sum of the single diffractive dissociation and elastic cross sections from the model and from experiments}\label{R_tab}
\end{center}
\end{table}
As can be seen our model predicts that the diffractive dissociation cross section grows more slowly with increasing energy than the elastic cross section. This is in qualitative agreement with experiment, where an even slower rise of $\sigma_{\mbox{\scriptsize{sd}}}$ compared to $\sigma_{\mbox{\scriptsize{el}}}$ is observed.

\begin{figure}[h]
  \unitlength1mm
  \begin{center}
    \begin{picture}(80,80)
      \put(0,0){\epsfig{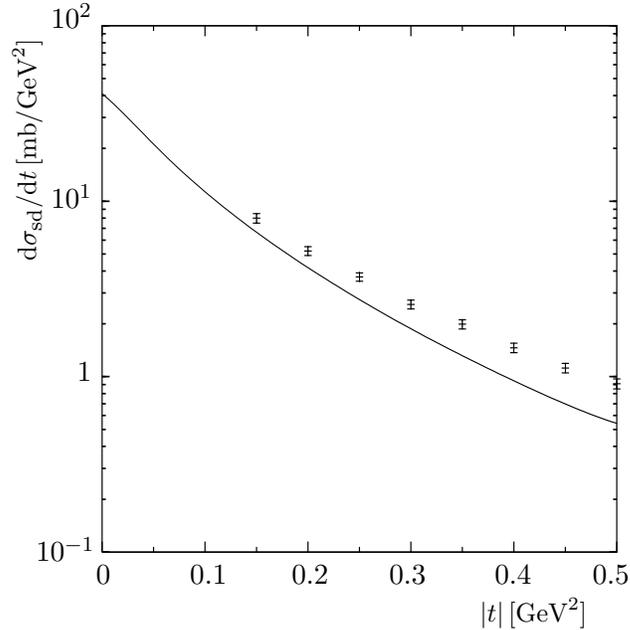}}
      \small
      \put(7,2){$0$}
      \put(19.5,2){$0.1$}
      \put(33,2){$0.2$}
      \put(46,2){$0.3$}
      \put(60,2){$0.4$}
      \put(73,2){$0.5$}
      \put(-1,5){$10^{-1}$}
      \put(4.5,29){$1$}
      \put(1,52){$10^1$}
      \put(1,76){$10^2$}
      \footnotesize
      \put(58,-3){$|t| \, [\mbox{GeV}^2]$}
      \put(-1,45){\begin{rotate}{90} $\mbox{d}\sigma_{\mbox{\scriptsize{sd}}} / \mbox{d}t \, [\mbox{mb}/\mbox{GeV}^2]$ \end{rotate}}
    \end{picture}
  \end{center}
  \vspace*{-0.3cm}
  \caption{{\em The differential diffractive cross section} $\mbox{d}\sigma_{\mbox{\scriptsize{sd}}} / \mbox{d}t \, [\mbox{mb}/\mbox{GeV}^2]$ {\em at} $\sqrt{s}=23.5 \mbox{ GeV}$ {\em compared to the experimental data from}~\cite{Albrow}}\label{dsig_pp_diss_fig}
\end{figure}
The result for the differential cross section of the single diffractive dissociation is shown in Fig.~\ref{dsig_pp_diss_fig}. The calculated curve describes the slope of the diffractive reaction quite well even for larger values of $|t|$. Therefore the agreement with the experiment is reasonably good within the $|t|$-range considered here.
This could however be partly accidental. In processes where the proton breaks up, the exchange of hard and semihard gluons will play an important role. This exchange is not described by our model which is an approximation for the infrared behaviour of QCD.
We have seen in elastic scattering that the expansion method overestimates the cross section for values of $|t|$ larger than $0.2 \, \mathrm{GeV}^2$, see Fig.~\ref{dsigdt_fig}, and this could simulate the expected contribution of hard or semihard gluon exchange in the diffractive dissociation reactions. We stress however that the fast decrease of the single diffractive dissociation cross section $\mathrm{d}\sigma_{\mathrm{sd}} / \mathrm{d}t$ for values $|t| \lesssim 0.2 \, \mathrm{GeV}^2$ is a firm prediction of our model.

To check the validity of our description of the diffractive final state by a free quark-diquark pair using plane waves, we now apply the second method, which describes the diffractive final state $X$ through a sum of wave functions of excited states of a two-dimensional harmonic oscillator, as explained above in section~\ref{scat_amp_hh_sect}.
In this description, the final state phase space is two-dimensional as in the case of elastic scattering and the differential cross section is given by
\begin{equation}
  \frac{\mbox{d}\sigma_{\mbox{\scriptsize{sd}}}}{\mbox{d}t}=\frac{1}{16 \pi} \frac{1}{s^2} \sum_{(n,m) \ne (0,0)}\left| \mathcal{T}_{fi} \right|^2
\end{equation}
with $\mathcal{T}_{fi}$ from~(\ref{J_osc}). The sum runs over all even $n$ for the reasons given in sect.~\ref{eval_hh_sect}, the associated quantum number $m$ runs over $m=-n,-(n-2),\ldots,n-2,n$. The numerical analysis shows that both calculations are in very good agreement to each other and that summing up the contributions from values of $n \le 6$ already gives $\approx 98\%$ of the result using plane waves.\newline\newline
In the following we will consider the mass spectrum $\mbox{d}^2 \sigma_{\mbox{\scriptsize{sd}}} / (\mbox{d} \xi \mbox{d}t)$ of the single diffractive dissociation reaction at $\sqrt{s}=23.5 \mbox{ GeV}$ for $t=-0.0525 \mbox{ GeV}^2$, where $\xi$ is the squared mass of the diffractive final state divided by $s$. In our ansatz with plane wave final states $\xi$ then is given by
\begin{equation}\label{xi}
  \xi:=\frac{M_X^2}{s}=\frac{{\bf \Delta}_{4T}^2+(1-z^\prime)m_q^2+z^\prime m_{\bar q}^2}{z^\prime(1-z^\prime)s}.
\end{equation}
Here $m_q$ and $m_{\bar q}$ are the masses of the quark and the diquark which describe the excited proton state. To take into account thresholds the mass for the quark has been chosen to be 330 MeV and for the diquark 660 MeV so that the sum roughly gives the proton mass. Going back to~(\ref{J_wave}) we recognise that we now can no longer replace the Gaussian shaped longitudinal momentum distribution in the wave function~(\ref{wave_funct}) for the hadron $h_2$ which breaks up by a delta function centred around 1/2, as we have done in the calculations before, because $z^\prime$ determines the value of $\xi$ in~(\ref{xi}). This was different for the calculation of $\mbox{d}\sigma_{\mbox{\scriptsize{sd}}} / \mbox{d}t$, where we performed an integration over the full range of ${\bf \Delta}_{4T}$ in phase space and were not interested in any particular value of $\xi$.
As a consequence of the introduction of quark masses the integration over $z^\prime$ now does not run from 0 to 1, but the integration limits are given by
\begin{equation}\label{xi_range}
  z^\prime_{0/1}=\frac{1}{2}-\frac{m_q^2-m_{\bar q}^2}{2 \xi s} \mp \sqrt{\frac{1}{4}-\frac{m_q^2+m_{\bar q}^2}{2 \xi s}+\left(\frac{m_q^2-m_{\bar q}^2}{2 \xi s}\right)^2}.
\end{equation}
This ensures that the mass spectrum starts at $M_X^2=M_p^2$ where $M_p$ is the proton mass. Our plane wave description of the diffractive final state of course also includes elastic scattering. To compare with experimental results on diffractive dissociation we have to subtract the elastic contribution.
To do so we argue as follows: to obtain the elastic contribution, we integrate $\mbox{d}^2\sigma_{\mbox{\scriptsize{sd}}} / (\mbox{d} \xi \mbox{d}t)$ over $\xi$ from $\xi_0 = M_p^2/s$ up to $\xi_1$. We then determine $\xi_1$ in such a way that the integral gives the value of the elastic differential cross section $\mathrm{d}\sigma_{\mathrm{el}} / \mathrm{d} t$. We now interprete the mass spectrum as consisting of the elastic part, which lies between $\xi_0$ and $\xi_1$ and the dissociation part, which starts at $\xi_1$. This procedure allows us to separate the elastic and the dissociation contributions.

The result of the calculation is shown in Fig.~\ref{d2sig_fig} for $t=-0.0525 \, \mathrm{GeV}^2$ and the c.m. energy $\sqrt{s}=23.5 \mbox{ GeV}$ together with the data points from~\cite{Armitage}, where we determined $\xi_1$ to be given by $\xi_1= 1.63 \, \mathrm{GeV}^2 / s \approx 0.003$.
The experimental values are smeared out over a certain range of values for $\xi$ because of the detector mass resolution function. This explains also the data for the unphysical negative $\xi$-values. As a consequence the large peak of the diffractive mass spectrum is much more pronounced in our calculation and the experimental distribution is flatter around that peak. To compare directly with the experiment, we would have to fold our results with the mass resolution function of the detector used in the experiments~\cite{Armitage}, but unfortunately, this resolution function can no longer be reconstructed~\cite{Donnachie}. We note that our model should give reliable results for small $\xi$. Indeed, for large values of $\xi$ the model seems to underestimate the data considerably. But for this $\xi$ region we expect, for instance, that our purely nonperturbative treatment of the scattering must be supplemented by hard gluon radiation which should lead to high invariant masses for the diffractively excited state. Furthermore our calculation treats the final state as a quark-diquark pair and therefore here no confinement effects are included.

\subsection{Proton-pion scattering}\label{results_ppi_sect}
We present calculations for the reaction $p \, \pi^\pm \to p \, X$. Of course, the vacuum parameters $G_2, a, \kappa$ stay the same but we still have to fix the pion extension parameters $S_\pi$ and $z_\pi$ in~(\ref{wave_funct}). Proceeding as in the case of proton-proton scattering we find for the parameters $S_\pi=0.60 \, \mathrm{fm}$ and $z_\pi=0.5$ at $\sqrt{s}=19.5 \mbox{ GeV}$.

Again we will first take a look at elastic scattering. For a c.m. energy of $\sqrt{s}=19.5 \mbox{ GeV}$ we find for the integrated elastic cross sections $\sigma_{\mbox{\scriptsize{el}}}=3.12 \, \mathrm{mb}$ compared to an experimental value of $\sigma_{\mbox{\scriptsize{el}}}=3.30 \pm 0.11 \, \mathrm{mb}$~\cite{Adamus}. Fitting our result for the differential cross section by $\mathrm{d}\sigma_{\mbox{\scriptsize{el}}} / \mathrm{d}t = A \exp b \, t$ we find $b=8.7 \pm 0.3 \, \mathrm{GeV}^{-2}$ compared to the experimentally measured value of $b=7.9 \pm 0.2 \, \mathrm{GeV}^{-2}$ for $\pi^+ \, p$ and $b=8.4 \pm 0.1 \, \mathrm{GeV}^{-2}$ for $\pi^- \, p$ scattering, respectively~\cite{ppi_elast_s20}. 

Moving on to the reaction where the pion breaks up diffractively, we calculate $\sigma_{\mbox{\scriptsize{sd}}}$ and the $R$-value, which we define as in the case of proton-proton scattering.
We find $\sigma_{\mbox{\scriptsize{sd}}}=1.99 \, \mathrm{mb}$ and $R=0.39$ where the experimental values are $\sigma_{\mbox{\scriptsize{sd}}}=1.90 \pm 0.2 \, \mathrm{mb}$ and $R=0.37 \pm 0.03$~\cite{Adamus2} which is quite good agreement.

\subsection{Photo- and electroproduction of $\rho^0$ mesons}
%
In this section we present the results for photo- and electroproduction of $\rho^0$ mesons proceeding exactly as described above.
Since several investigations on the reaction
$ \gamma^{(*)} p \rightarrow \rho^0 p $
in the approach discussed here have been published \cite{DGKP,KDP99} we concentrate on the single diffractive dissociation reaction
$ \gamma^{(*)} p \rightarrow \rho^0 X $,
where $X$ does not include the proton.\\
The energy dependence introduced in chapter 4 cannot explain the strong dependence which has been observed at HERA for photons with high
$Q^2$.
We therefore
do the calculations with the parameters fixed for $\sqrt{s}=23.5\mbox{ GeV}$ and
concentrate on ratios of cross sections
, which should be more or less energy independent.
For photoproduction we find
\begin{equation}
 \frac{\sigma_{\mathrm{sd}}(\gamma p \rightarrow \rho^0 X)}{\sigma_{\mathrm{el}}(\gamma p \rightarrow \rho^0 p)}=0.6 \;,
\end{equation}
which agrees within the errors with the experimental value 
$ \sigma_{\mathrm{sd}}/\sigma_{\mathrm{el}} = 0.5^{+0.4}_{-0.16}$ from \cite{Breit},
which was measured in the range $50\mbox{ GeV}^2 \le \sqrt{s} \le 100\mbox{ GeV}^2$.
In our model the ratio of single diffractive dissociation to elastic $\rho^0$ production depends very weakly on $Q^2$.
For the range $7\mbox{ GeV}^2 \le Q^2 \le 35\mbox{ GeV}^2$ we find
\begin{equation}
 \frac{\sigma_{\mathrm{sd}}(\gamma^* p \rightarrow \rho^0 X)}{\sigma_{\mathrm{el}}(\gamma^* p \rightarrow \rho^0 p)}=0.54 \;,
\end{equation}
which also agrees with the experimental value $ \sigma_{\mathrm{sd}}/\sigma_{\mathrm{el}} = 0.65 \pm 0.24$ from \cite{Adlo}
(here $60\mbox{ GeV}^2 \le \sqrt{s} \le 180\mbox{ GeV}^2$).
The $Q^2$ dependence of our model calculations can be fitted by $\sigma_{\mathrm{sd}}(Q^2) \propto (Q/Q_0)^{-5.0}$
compared with $\sigma_{\mathrm{sd}}(Q^2) \propto (Q/Q_0)^{-5.8 \pm 1.9}$ from experiment.
The polarisation ratio of the outgoing $\rho^0$ mesons has been calculated in the range $7\mbox{ GeV}^2 \le Q^2 \le 35\mbox{ GeV}^2$:
\begin{equation}
 \frac{\sigma_\mathrm{L,sd}}{\sigma_\mathrm{L,sd}+\sigma_\mathrm{T,sd}} = 0.71 \;.
\end{equation}
This has to be compared with the experimental result
$\sigma_\mathrm{L,sd}/(\sigma_\mathrm{L,sd}+\sigma_\mathrm{T,sd}) = 0.79 \pm 0.15$.\\

\bigskip
\section{Form factors in the model}\label{ff_sect}

In this chapter we will study form factors within our model. We do not intend to perform a precision calculation of form factors but we will apply the calculation to extract values for the width of the longitudinal momentum distributions of the proton and the pion, $z_p$ and $z_\pi$, respectively, by fitting our results to experimental data.

\subsection{The electromagnetic form factors of the proton}
The coupling of the electromagnetic current to the proton can be described by
\begin{equation}\label{ff_matrix}
  \langle p(P^\prime,s^\prime) | j^\mu(0) | p(P,s) \rangle = \bar u_{s^\prime}(P^\prime) \left[ F_{1p}(Q^2) \gamma^\mu + \frac{i \sigma^{\mu\nu} q_\nu}{2M_p} F_{2p}(Q^2) \right] u_s(P),
\end{equation}
where the momentum transfer is $q=P^\prime-P$, $Q^2=-q^2$, $M_p$ is the proton mass and $F_{1p}, F_{2p}$ are the Dirac and Pauli form factor of the proton, respectively.
Now we choose a coordinate system such that $q$ is purely transverse:
\begin{eqnarray}
  P^\mu &=& \frac{1}{2}P_+ n_+^\mu + \frac{1}{2} P_- n_-^\mu - \frac{1}{2} q^\mu, \nonumber \\
  P^{\prime \mu} &=& \frac{1}{2}P_+ n_+^\mu + \frac{1}{2} P_- n_-^\mu + \frac{1}{2} q^\mu, \nonumber \\
  q &=&\left(\begin{array}{c}0\\{\bf q}_T\\0\end{array}\right), \qquad n_\pm=\left(\begin{array}{c}1\\0\\0\\ \pm 1\end{array}\right), \nonumber \\
  P_- &=& (\frac{1}{4} {\bf q}_T^2 + M_p^2 )/P_+. \label{coor_sys}
\end{eqnarray}
In the high energy limit, $P_+ \to \infty$, we get for the matrix element~(\ref{ff_matrix}) (see~\cite{DY70})
\begin{equation}\label{strommatrix}
  \langle p(P^\prime,s^\prime) | j^\mu(0) | p(P,s) \rangle = P_+ n_+^\mu \, \chi_{s^\prime}^\dagger \left[ F_{1p}(Q^2) - \frac{\sigma^3 {\bf q}_T \cdot \boldsymbol{\sigma}}{2 M_p} F_{2p}(Q^2) \right] \chi_s + \mathcal{O}(1),
\end{equation}
where $\chi_s, \chi_{s^\prime}$ are the Pauli two-component spinors. $F_{1p}$ multiplies the spin-non-flip part, $F_{2p}$ the spin-flip part of the matrix element.
Calculating the spin average of this expression leads to
\begin{equation}
  \frac{1}{2}\sum_s \langle p(P^\prime,s) | j^\mu(x) | p(P,s) \rangle = P_+ n_+^\mu F_{1p}(Q^2) + \mathcal{O}(1).
\end{equation}
We describe the calculation of the Dirac form factor of the proton in the framework of our model in Appendix~\ref{App_B_sect}. In the following we consider the matrix element of the third component of the isospin current $j_3^\mu$. Its matrix element between proton states is as in~(\ref{ff_matrix}),(\ref{strommatrix}), with $F_{ip}$ replaced by $F_{iv}$, related to the form factors of proton and neutron by
\begin{equation}\label{ff_pn}
  F_{iv}=\frac{1}{2}\left(F_{ip}(Q^2)-F_{in}(Q^2)\right) \qquad (i=1,2).
\end{equation}
With the wave functions~(\ref{wave_funct}) we obtain
\begin{eqnarray} \label{ff}
  F_{1v}(Q^2) &=& \frac{1}{2 I_p} \int_0^1 \mbox{d}z \, 2 z (1-z) \, e^{-(z-\frac{1}{2})^2 / 2 z_p^2} \, e^{-\frac{z^2}{2} S_p^2 Q^2},
\end{eqnarray}
where $I_p$ is the normalisation factor~(\ref{norm_z}).
For this calculation we need only the expectation value of one Wegner-Wilson loop. The expectation value over one single loop is 1 in both the matrix cumulant method and the expansion method. Thus, in our model the form factor is just the Fourier transform of the squared wave function.

\begin{figure}[h]
  \unitlength1mm
  \begin{center}
    \begin{picture}(120,90)
      \put(0,0){\epsfysize=90mm \epsffile{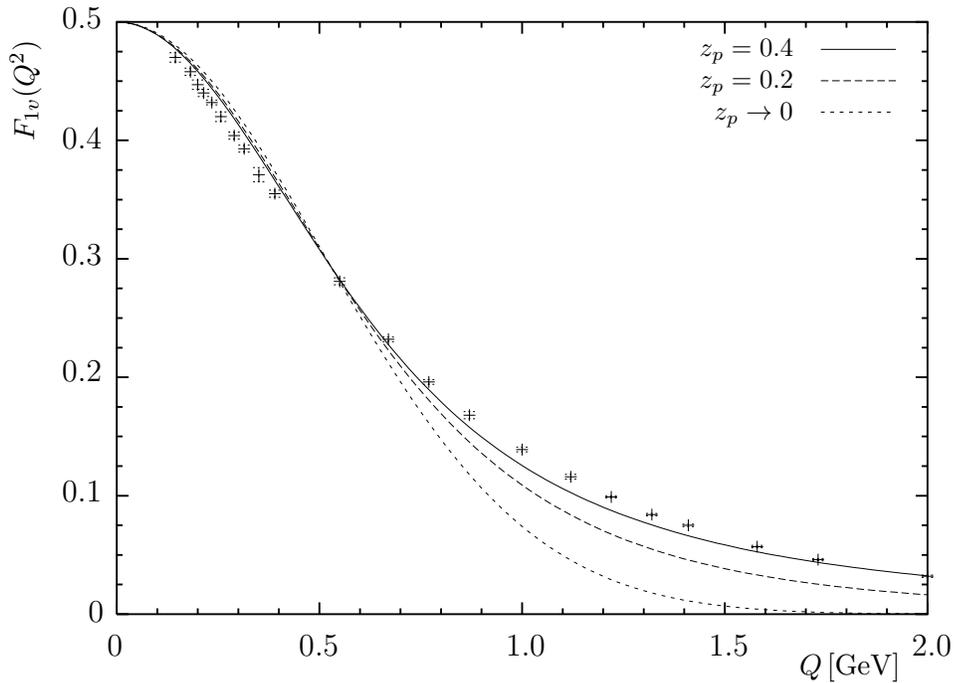}}
      \put(11,1.5){$0$}
      \put(36.5,1.5){$0.5$}
      \put(64,1.5){$1.0$}
      \put(91,1.5){$1.5$}
      \put(118,1.5){$2.0$}
      \put(8.5,6){$0$}
      \put(5,22){$0.1$}
      \put(5,38){$0.2$}
      \put(5,53.5){$0.3$}
      \put(5,69){$0.4$}
      \put(5,85){$0.5$}
      \small
      \put(103,-1){$Q \, [\mbox{GeV}]$}
      \put(0,70){
        \begin{rotate}{90}{$F_{1v}(Q^2)$}\end{rotate}
      }
      \footnotesize
      \put(88,77){$\begin{array}{r} z_p=0.4 \\ z_p=0.2 \\ z_p \rightarrow 0 \end{array}$}
    \end{picture}
  \end{center}
  \vspace*{-0.5cm}
  \caption{{\em The isovector form factor of the proton for $S_p=0.77 \, \mathrm{fm}$ and different values of $z_p$ compared to the experimental data from}~\cite{Borkowski,ff_p_daten,Schmieden}}\label{ff_p_fig}
\end{figure}

We will now use~(\ref{ff}) to determine $z_p$ and $S_p$. It turns out that in the range $0 \le Q \le 0.5 \, \mathrm{GeV}$ the form factor depends sensitively on $S_p$ but only weakly on $z_p$. From a fit to experiment in this region we obtain $S_p=0.77 \, \mathrm{fm}$.
With $S_p$ fixed to this value we show in Fig.~\ref{ff_p_fig} our result~(\ref{ff}) for $F_{1v}$ for different values of $z_p$. The experimental values have been calculated from the experimental data for $G_{Ep}$ and $G_{Mp}$ from~\cite{Borkowski,ff_p_daten} and a fit of the experimental data on $G_{En}$ and $G_{Mn}$~\cite{Schmieden} according to~(\ref{ff_pn}) and the relation between the Dirac ($F_{1 p,n}$) and the electric ($G_{E p,n}$) and magnetic ($G_{M p,n}$) form factor of the proton and neutron, respectively:
\begin{equation}\label{ff_DEM}
  F_{1 N}(Q^2) = \frac{G_{E N}(Q^2) + \tau G_{M N}(Q^2)}{1 + \tau}, \qquad \tau = \frac{Q^2}{4M_{N}^2} \qquad (N=p,n).
\end{equation}
The best fit is found for $z_p=0.4$. As can be seen from Fig.~\ref{ff_p_fig}, $z_p$, which fixes the width of the longitudinal momentum distribution of the constituents, plays no important role for $Q \lesssim 0.5 \, \mathrm{GeV}$. For larger values of $Q$ however, our fit is substantially improved when using a Gaussian shaped $z$-dependence instead of a delta-function centred around $z=1/2$, which is equivalent to $z_p \to 0$.

It has to be noted, that the proton extension parameter $S_p$ obtained from~(\ref{ff}) is not, and need not be, the same as the one used in the hadronic scattering processes in the previous chapters. Whereas the hadronic extension parameter has been allowed to be energy dependent (see~(\ref{S2_eikonal}),(\ref{S2_chi2})) to account for the rise of $\sigma_{\mathrm{tot}}$ with $\sqrt{s}$, the extension parameter connected with the form factor has a fixed value for all energies as the form factor itself is energy independent and is related to the electromagnetic radius of the proton as follows.
Using the definitions
\begin{eqnarray} \label{dFdQ2}
  \langle r^2 \rangle^p &=& -6 \left.\frac{\mbox{d}G_{Ep}(Q^2)}{\mbox{d}Q^2}\right|_{Q^2=0}, \nonumber \\
  r^p_{\mathrm{em}} &=& \sqrt{\langle r^2 \rangle^p},
\end{eqnarray}
relations~(\ref{ff_pn}),(\ref{ff_DEM}) and the experimental value
\begin{equation}
  \left.\frac{\mbox{d}G_{En}(Q^2)}{\mbox{d}Q^2}\right|_{Q^2=0}=0.019 \, \mathrm{fm}^2
\end{equation}
from thermal-neutron-electron scattering~\cite{Leeb}, we get from our model
\begin{equation}\label{r_em}
  r^p_{\mathrm{em}}=0.81 \, \mathrm{fm}.
\end{equation}
This coincides with the value one obtains for the proton electromagnetic radius when describing the electric form factor of the proton by the dipole parametrisation~\cite{Borkowski}, which also results in $r^p_{\mathrm{em}}=0.81 \, \mathrm{fm}$.
From scattering experiments one finds $r^p_{\mathrm{em}}=0.88 \pm 0.03 \, \mathrm{fm}$ or $r^p_{\mathrm{em}}=0.92 \pm 0.03 \, \mathrm{fm}$, depending on which fit is used for the experimental data on $G_{Ep}(Q^2)$ for small $Q^2$~\cite{Borkowski}. The Lamb shift measurements~\cite{Udem} give $r^p_{\mathrm{em}}=0.890 \pm 0.014 \, \mathrm{fm}$.
Thus our result~(\ref{r_em}), as well as the one calculated from the dipole parametrisation, is smaller than the experimental value for $r^p_{\mathrm{em}}$. Our calculation as well as the dipole fit describe the data~\cite{ff_p_daten} for $G_{Ep}$ rather well for $Q \gtrsim 0.4 \, \mathrm{GeV}$. But for smaller $Q$ the data~\cite{Borkowski} indicate a rapid change in the slope $\mathrm{d}G_{Ep}(Q^2)/\mathrm{d}Q^2$ which is described neither by our model nor by the dipole parametrisation. Such an ``anomalous'' behaviour of $G_{Ep}$ and $G_{En}$ for small $Q^2$ has been related to QCD vacuum effects in \cite{ON_Vac}. 

\subsection{The electromagnetic form factor of the pion}
For the charged pions $\pi^\pm$ the matrix element of the electromagnetic and the third component of the isospin current are equal. Choosing again the coordinate system as in~(\ref{coor_sys}) with $M_p$ replaced by $m_\pi$ we get
\begin{equation}
  \langle \pi^+(P^\prime) | j^\mu(0) | \pi^+(P) \rangle = (P_+ n_+^\mu + P_- n_-^\mu) F_\pi(Q^2).
\end{equation}
Here the matrix element can be expressed by only one form factor $F_\pi$.
The calculation of this matrix element in our model leads to
\begin{eqnarray}\label{ff_pi}
  F_\pi(Q^2)=\frac{1}{I_\pi} \int_0^1 \mbox{d}z \, 2 z (1-z) \, e^{-(z-\frac{1}{2})^2 / 2 z_\pi^2} \, e^{-\frac{z^2}{2} S_\pi^2 Q^2}.
\end{eqnarray}
\begin{figure}[H]
  \unitlength1mm
  \begin{center}
    \begin{picture}(120,80)
      \put(0,0){\epsfysize=90mm \epsffile{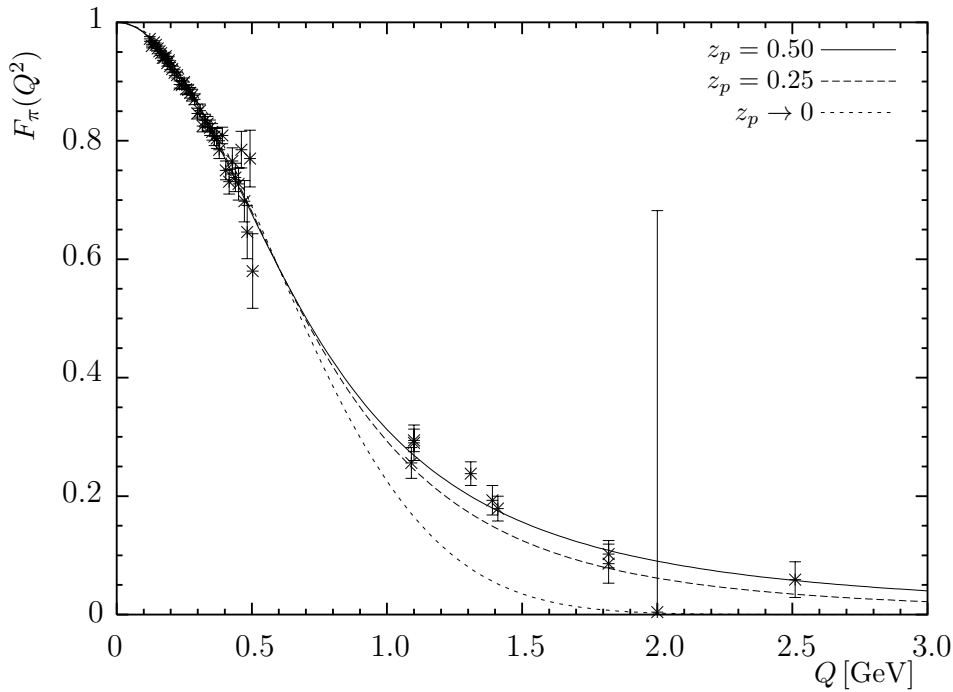}}
      \put(11,1.5){$0$}
      \put(28,1.5){$0.5$}
      \put(46,1.5){$1.0$}
      \put(64,1.5){$1.5$}
      \put(82.5,1.5){$2.0$}
      \put(100,1.5){$2.5$}
      \put(118,1.5){$3.0$}
      \put(8.5,6){$0$}
      \put(5,22){$0.2$}
      \put(5,38){$0.4$}
      \put(5,53.5){$0.6$}
      \put(5,69){$0.8$}
      \put(8.5,85){$1$}
      \small
      \put(105,-2){$Q \, [\mbox{GeV}]$}
      \put(0,70){
        \begin{rotate}{90}{$F_\pi(Q^2)$}\end{rotate}
      }
      \footnotesize
      \put(89,77){$\begin{array}{r} z_p=0.50 \\ z_p=0.25 \\ z_p \rightarrow 0 \end{array}$}
    \end{picture}
  \end{center}
  \vspace*{-0.5cm}
  \caption{{\em The pion form factor for $S_\pi=0.68 \, \mathrm{fm}$ and different values of $z_\pi$ compared to the experimental data from}~\cite{pion_daten}}\label{ff_pi_fig}
\end{figure}
We compare~(\ref{ff_pi}) to experimental data for $F_\pi$ from~\cite{pion_daten} in Fig.~\ref{ff_pi_fig}.
As for the proton the transverse extension parameter $S_\pi$ can be fitted in the range $0 \le Q \le 0.5 \, \mathrm{GeV}$ with the result $S_\pi=0.68 \, \mathrm{fm}$. Using the analogue of relation~(\ref{dFdQ2}) for the pion, this value gives an electromagnetic radius $r^\pi_{\mathrm{em}}=0.64 \, \mathrm{fm}$, which is consistent with the experimental value $r^\pi_{\mathrm{em}}=0.663 \pm 0.006 \, \mathrm{fm}$~\cite{pion_daten}.
For values $Q \gtrsim 0.5 \, \mathrm{GeV}$ our fit becomes sensitive to the width of the longitudinal momentum distribution of the constituents. For the pion, the best fit for the width of this distribution is given by $z_\pi=0.5$. The broader distribution compared to the proton is related to the smaller mass of the pion, which is in agreement with the parametrisation of the hadron wave functions in~\cite{BWS}.

\bigskip
\section{Conclusions}\label{concl_sect}

In this work we have calculated total and differential cross sections for elastic and inelastic diffractive scattering at high c.m energies and small momentum transfer.
In our model we start from a microscopic description of the scattering of quark-antiquark and quark-diquark wave packets and use functional integral methods to get expressions for the scattering amplitudes. The correlation functions of light-like Wegner-Wilson loops governing these amplitudes are evaluated in the framework of the model of the stochastic vacuum~\cite{MSVmink,DFK,MSV,DRio}.
The hadron-hadron scattering amplitudes are obtained by multiplying the parton scattering amplitudes with suitable hadronic wave functions~\cite{BWS}, or the photon wave function~\cite{DGKP} in the case of photo- and electroproduction.
Both a matrix cumulant expansion for the correlation function of two Wegner-Wilson loops as developed in~\cite{Berger} and an expansion method~\cite{DFK,DGKP} are used.

The free parameters of our model are those of the model of the stochastic vacuum: $G_2$, $a$ and $\kappa$, and the ones in the wave functions: $S_{h_i}$ and $z_{h_i}$, determining the width of the transverse and longitudinal momentum distributions of the constituents of the hadrons, respectively. 
These parameters have been determined in previous work~\cite{DFK,Berger} on elastic scattering.
The extension parameters $S_{h_i}$ are allowed to depend on the c.m. energy according to~(\ref{S2_eikonal}) and~(\ref{S2_chi2}) respectively. In this sense different hadrons are characterised through their radii, which come out close to the corresponding electromagnetic radii of the hadrons for energies $\sqrt{s} \approx 20 \mbox{ GeV}$.
The values for $z_{h_i}$ are obtained from a calculation of form factors in our model. Our result for the isovector Dirac form factor of the proton and the electromagnetic form factor of the pion, as well as the electromagnetic radii extracted from them, compare reasonably well to experimental data.

With all parameters fixed, integrated and differential cross sections are calculated and compared to experimental results~\cite{pp_elast_s23,Armitage,Bernard,Abe,Albrow}.
The calculated integrated elastic cross sections agree with the experimental values within the numerical and experimental errors for a wide range of c.m. energies starting at about $\sqrt{s}=20 \mbox{ GeV}$ up to the Tevatron energy $\sqrt{s}=1800 \mbox{ GeV}$. Our model does not distinguish between $pp$ and $p \bar p$ scattering or $p \pi^+$ and $p \pi^-$ scattering, respectively. In the approximation we use here we have $C=+1$ exchange only.

Furthermore the rise of the integrated cross sections in inelastic diffractive reactions as a function of $\sqrt{s}$ is calculated. Our calculated ratio $\sigma_{\mathrm{sd}} / (\sigma_{\mathrm{el}} + \sigma_{\mathrm{sd}})$ is in rough agreement with experiment. The experimentally observed behaviour that the diffractive dissociation part of the cross section increases more slowly with $\sqrt{s}$ than the elastic one is reproduced qualitatively in our calculation.

For photo- and electroproduction of $\rho^0$ mesons we restrict ourselves to ratios of integrated cross sections. The ratio of single diffractive dissociation to elastic cross sections is found to be in the range $0.5 \ldots 0.6$ for $Q^2 \le 35 \mbox{ GeV}^2$, consistent with the experimental results. The $Q^2$-dependence of the integrated single diffractive dissociation cross section as well as the polarisation ratio of the produced $\rho^0$ mesons also agree with the experimental values. 

To summarise, our model is quite well suited to describe inelastic diffractive hadronic reactions at high c.m. energies ($\sqrt{s} \gtrsim 20 \mbox{ GeV}$) and small momentum transfer. Further progress could be made when including higher cumulant terms in~(\ref{eikonal_result}) which would contribute to both $C=+1$ and $C=-1$ exchange.

Another area where our model can be applied is double diffractive dissociation, where $C=-1$ contributions (odderon exchange) occur already in our present approximation. This will be the topic of a following investigation. The upcoming experiments at RHIC will be a rich source for new experimental data for both single and double diffractive dissociation in hadronic reactions at high c.m. energies. Therefore the study of inelastic diffractive scattering will remain an interesting and instructive field of work, where effects of nonperturbative QCD can be studied.

\bigskip
\section*{Acknowledgements}
We would like to thank E.R. Berger, A. Donnachie, A. Hebecker, G. Kulzinger, D.I. Melikhov, H.J. Pirner, A. Rauscher and M. R\"uter for valuable discussions.\\
\bigskip

\begin{appendix}
\section{Definition of ${\bf \Delta}_T$}\label{App_A_sect}

The light-cone momenta of the quark and the antiquark (or diquark) of the diffractive final state $X$ are given by
\begin{eqnarray}
\begin{array}{lclclcl}
p_-&=&z^\prime P_-, &\quad& p^\prime_-&=&(1-z^\prime) P_-, \\
{\bf p}_T &=& z^\prime {\bf P}_T+{\bf \Delta}_T, &\quad& {\bf p}^\prime_T&=&(1-z^\prime) {\bf P}_T-{\bf \Delta}_T,
\end{array}
\end{eqnarray}
where $P_-$ is the longitudinal and ${\bf P}_T$ the transverse momentum of $X$, which moves in negative $x^3$ direction, therefore $P_+ \approx 0$. The longitudinal momentum fraction carried by the quark is $z^\prime$. Lorentz invariance requires $z^\prime$ to appear also in the transverse components of ${\bf p}_T$ and ${\bf p}^\prime_T$ as defined above. The relative transverse momentum between the quark and the antiquark (or diquark) is given by
\begin{equation}
  {\bf \Delta}_T=\frac{{\bf p}_T-{\bf p}^\prime_T}{2}+\left(\frac{1}{2}-z^\prime\right){\bf P}_T.
\end{equation}
\section{Calculation of the form factor}\label{App_B_sect}

Starting point for the form factor calculation is the matrix element of the third component of the isospin current at $x=0$
\begin{eqnarray}\label{current_matrix}
  J_3^\mu \equiv \langle h_3(P^\prime) | j_3^\mu(0) | h_1(P) \rangle
\end{eqnarray}
with 
\begin{equation}
  j_3^\mu(x) = \sum_\psi \bar \psi(x) \gamma^\mu \left(\frac{1}{2}\tau^3\right) \psi(x)
\end{equation}
Here $\sum_\psi$ denotes the sum over quark fields $u,d$ and $\tau^3$ is the third Pauli isospin matrix.
The hadrons $h_1,h_3$ are supposed to move in positive $x^3$-direction with $P_+=P_+^\prime \to \infty$ (see~\ref{coor_sys}). In analogy to the description of hadron-hadron scattering in~\cite{DFK,Nachtmannreport} we therefore denote the incoming hadron by $h_1$ and the outgoing hadron by $h_3$.
The steps required to compute the form factor from this expression are completely analogous to those discussed in~\cite{DFK,Nachtmannreport} that lead to the $\mathcal{T}$-matrix element~(\ref{scat_amp}), with the difference being that there are now additional contractions between the quarks and diquarks (or antiquarks in the case of mesons) of the hadrons $h_1,h_3$ and the quark fields of the current $j_3^\mu$ when applying the LSZ reduction formalism. By considering the isospin current we ensure that contributions which contain subdiagrams arising from contractions between the quark fields of the current drop out because they are proportional to $\mathrm{Tr} \, \tau^3=0$.
We describe now the form factor calculation for the $\pi^+$ meson, modeled as $u \bar d$ wave packet.

\begin{figure}[h]
\vspace*{1cm}
\begin{center}\unitlength1mm
\begin{picture}(200,50)
\put(30,0){\epsfysize=50mm \epsffile{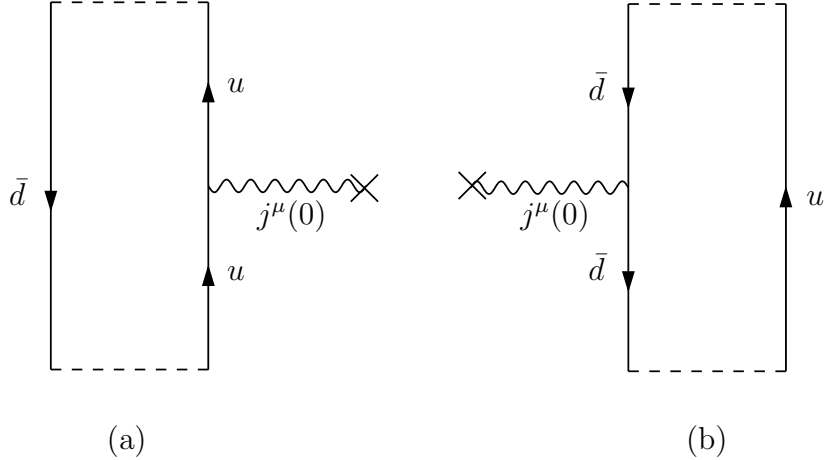}}
\put(25,22.5){$\bar d$}
\put(54,12.5){$u$}
\put(54,37.5){$u$}
\put(58,20){$j^\mu(0)$}
\put(131,22.5){$u$}
\put(102,12.5){$\bar d$}
\put(102,36.5){$\bar d$}
\put(93,20){$j^\mu(0)$}
\put(38,-10){(a)}
\put(115,-10){(b)}
\end{picture}
\end{center}
\vspace*{0.7cm}
\caption{\em The two contributions to the matrix element $\langle u \bar d | j_3^\mu | u \bar d \rangle$}\label{ff_pic}
\end{figure}

Using the same notation as in~\cite{Nachtmannreport} we obtain $J_3^\mu$~(\ref{current_matrix}) by first calculating the matrix element of $j_3^\mu$ between $q \bar q$ states and then folding with the wave functions of the wave packets.
\begin{eqnarray}
  J_3^\mu &=& \int \mbox{d}^2 \Delta_{3T} \int_0^1 \mbox{d}z_3 \, \frac{1}{(2 \pi)^{3/2}} \frac{1}{\sqrt{2}} \, \delta_{s_3,-s_3^\prime} \tilde \varphi_3^*(z_3,{\bf \Delta}_{3T}) \frac{1}{\sqrt{3}} \, \delta_{A_3 A_3^\prime} \nonumber \\
& &\int \mbox{d}^2 \Delta_{1T} \int_0^1 \mbox{d}z_1 \, \frac{1}{(2 \pi)^{3/2}} \frac{1}{\sqrt{2}} \, \delta_{s_1, -s_1^\prime} \tilde \varphi_1(z_1,{\bf \Delta}_{1T}) \frac{1}{\sqrt{3}} \, \delta_{A_1 A_1^\prime} \langle u \bar d | j_3^\mu | u \bar d \rangle \nonumber \\
 \langle u \bar d | j_3^\mu | u \bar d \rangle &\equiv& \quad \langle u(p_3,s_3,A_3) \bar d(p_3^\prime,s_3^\prime,A_3^\prime) | j_3^\mu(0) | u(p_1,s_1,A_1) \bar d(p_1^\prime,s_1^\prime,A_1^\prime) \rangle, \label{qq_matrix}
\end{eqnarray}
where $s_i, A_i$ are spin and colour indices, respectively and $\tilde \varphi_{1,3}$ are the Fourier transforms of the wave functions~(\ref{wave_funct})
\begin{equation}
  \tilde \varphi_i(z, {\bf \Delta}_T) = \frac{1}{2 \pi} \int \mathrm{d}^2 x_T \, e^{-i {\bf \Delta}_T \cdot {\bf x}_T } \varphi_i(z, {\bf x}_T).
\end{equation}
Applying the LSZ reduction formalism we can express the matrix element $\langle u \bar d | j_3^\mu | u \bar d \rangle$~(\ref{qq_matrix}) as an integral over the quark 6-point-function. We get here only two terms depicted graphically in Fig.~\ref{ff_pic} which are to be interpreted as follows. We consider a fixed gluon background. The quark and antiquark travel in this background and the current either hooks onto the quark line (Fig.~\ref{ff_pic}a) or the antiquark line (Fig.~\ref{ff_pic}b). The matrix element~(\ref{qq_matrix}) is then obtained by averaging over all gluon potentials with a measure given by the functional integral (see~\cite{Nachtmannreport} for the details). In the high energy limit for $u$ und $\bar d$ the scattering amplitudes in the fixed gluon background reduce to Wegner-Wilson line operators which are closed to a loop $\mathcal{W}_+$ by the meson wave functions. This is indicated by the dashed lines in Fig.~\ref{ff_pic}. Putting everything together we obtain
\begin{eqnarray}
  \langle \pi^+(P^\prime) | j_3^\mu(0) | \pi^+(P) \rangle &=& \frac{P_{1+} n_+^\mu}{2} \int_0^1 \mathrm{d}z \, \int \mathrm{d}^2x_T \, \varphi^*_3(z, {\bf x}_T) \varphi_1(z, {\bf x}_T) \nonumber \\
    & & \Biggl( e^{i(1-z) {\bf q}_T \cdot {\bf x}_T}  +  e^{-iz {\bf q}_T \cdot {\bf x}_T} \Biggr) \left\langle \mathcal{W}_+(\frac{1}{2}{\bf x}_T,{\bf x}_T) \right\rangle_G.
\end{eqnarray}

A straightforward calculation in the MSV gives that the expectation value of the correlation function of one Wegner-Wilson loop is equal to 1. By a shift in the integration variable the $\mathrm{d}^2 x_T$ integration can then be reduced to a Gaussian integral over the wave functions and we find as the final result for the matrix element~(\ref{current_matrix})
\begin{eqnarray}\label{piff}
  \langle \pi^+(P^\prime) | j_3^\mu(0) | \pi^+(P) \rangle = \frac{P_{1+} n_+^\mu} {I_\pi} \int_0^1 \mathrm{d}z \, 2z(1-z) \, e^{-(z-\frac{1}{2})^2/2z_h^2} \, e^{-\frac{z^2}{2} S^2_h {\bf q}_T^2}.
\end{eqnarray}

Let us now turn to the proton form factors.
In our simple ansatz the proton consists of a quark and a scalar diquark, which should be favoured over the vector diquark due to dynamical reasons~\cite{SSV94}.
The spin of the proton is then carried by the quark. This together with the spin conservation on the parton level has as consequence that, in our model, we get for the matrix element of $j_3^\mu$ between proton states an expression similar to~(\ref{piff}):
\begin{equation}
  \langle p(P^\prime,s^\prime) | j_3^\mu(0) | p(P,s) \rangle = P_+ n_+^\mu \chi_{s^\prime}^\dagger F_{1v}(Q^2) \chi_s
\end{equation}
with $F_{1v}(Q^2)$ given in~(\ref{ff}). Thus we get only a spin non flip and no spin flip contribution in the matrix element~(\ref{strommatrix}), that is, out model gives $F_{2v}(Q^2)=0$.
This is certainly not a very good approximation. But on the other hand the spin flip part in~(\ref{strommatrix}) is suppressed by $|{\bf q}_T| / (2 M_p)$ for ${\bf q}_T \to 0$. Thus the matrix element~(\ref{strommatrix}) is still reasonably described by the model for small enough  $|{\bf q}_T|$.

To summarise, we have outlined in this appendix a calculation of isovector form factors using the same methods as for the scattering processes. The results are in essence as in~\cite{DY70} taking our simple ansatz for the wave functions of the hadrons into account.

\end{appendix}

\end{document}